\documentclass[twocolumn]{aastex631}
\usepackage{natbib, graphicx}
\usepackage{amsmath, wrapfig, rotating}
\newcommand\HI{H\,{\sc i} }
\graphicspath{{./}{Figures/}}

\shorttitle{A Generalist ALFALFA BTFR}
\shortauthors{Ball et al.}

\begin{document}

\title{A Generalist, Automated ALFALFA Baryonic Tully-Fisher Relation}

\correspondingauthor{Catie J. Ball}
\email{cjb356@cornell.edu}

\author[0000-0002-1895-0528]{Catie J. Ball}
\affiliation{Cornell Center for Astrophysics and Planetary Science,
Space Sciences Building, Cornell University,
Ithaca, NY, 14853, USA}
\author[0000-0001-5334-5166]{Martha P. Haynes}
\affiliation{Cornell Center for Astrophysics and Planetary Science,
Space Sciences Building, Cornell University,
Ithaca, NY, 14853, USA}
\author[0000-0002-5434-4904]{Michael G. Jones}
\affiliation{Steward Observatory, University of Arizona, 933 North Cherry Avenue, Rm. N204, Tucson, AZ 85721-0065, USA}
\author[0000-0002-1605-0032]{Bo Peng}
\affiliation{Cornell Center for Astrophysics and Planetary Science,
Space Sciences Building, Cornell University,
Ithaca, NY, 14853, USA}
\author[0000-0002-3406-5502]{Adriana Durbala}
\affiliation{Department of Physics and Astronomy, University of Wisconsin -- Stevens Point, Stevens Point, WI 54481, USA}
\author[0000-0002-3144-2501]{Rebecca A. Koopmann}
\affiliation{Department of Physics and Astronomy, Union College, Schenectady, NY 12308, USA}
\author[0000-0003-3381-9795]{Joseph Ribaudo}
\affiliation{Department of Engineering and Physics, Providence College, Providence, RI 02918, USA}
\author[0000-0003-1664-6255]{Aileen A. O'Donoghue}
\affiliation{Department of Physics, St. Lawrence University, 23 Romoda Drive, Canton, NY 13617, USA}



\begin{abstract}
The Baryonic Tully-Fisher Relation (BTFR) has applications in galaxy evolution as a testbed for the galaxy-halo connection and in observational cosmology as a redshift-independent secondary distance indicator.
We use the 31,000+ galaxy ALFALFA sample -- which provides redshifts, velocity widths, and \HI content for a large number of gas-bearing galaxies in the local universe -- to fit and test an extensive local universe BTFR. This BTFR is designed to be as inclusive of ALFALFA and comparable samples as possible. Velocity widths measured via an automated method and $M_{b}$ proxies extracted from survey data can be uniformly and efficiently measured for other samples, giving this  analysis broad applicability. We also investigate the role of sample demographics in determining the best-fit relation. We find that the best-fit relations are changed significantly by changes to the sample mass range and to second order, mass sampling, gas fraction, different stellar mass and velocity width measurements. We use a subset of ALFALFA with demographics that reflect the full sample to measure a robust BTFR slope of $3.30\pm0.06$. We apply this relation and estimate source distances, finding general agreement with flow-model distances as well as average distance uncertainties of $\sim0.17$ dex for the full ALFALFA sample. We demonstrate the utility of these distance estimates by applying them to a sample of sources in the Virgo vicinity, recovering signatures of infall consistent with previous work.

\end{abstract}



\section{Introduction} \label{sec:intro}

The Baryonic Tully Fisher Relation (BTFR) is the tight empirical relation between the rotational velocity and baryonic mass of disk galaxies, extending the Tully-Fisher relation (TFR) between stellar luminosity (tracing stellar mass) and rotational velocity \citep{tully77} to lower mass galaxies with higher gas fractions \citep{mcgaugh00}. Fundamentally, the BTFR traces the relationship between dynamical and luminous mass in galaxies - in $\Lambda$CDM, this is the relationship between dark matter halos and the baryons which occupy them.  

The BTFR is thus a testbed for cosmology and galaxy evolution, as its slope, scatter, and normalization are set by this connection. Simple schematic arguments which assume the mass of a halo is defined by a density contrast threshold and constant baryon fraction predict a power law $M \propto V^3$. More robust treatments in $\Lambda$CDM can accommodate larger slopes ($\sim 3.4$, \citet{zaritsky14}), and semi-analytic models which incorporate abundance matching predict curvature in the BTFR as a consequence of the non-uniform baryon fraction \citep{trujillo-gomez11}. Some studies suggest that the BTFR may pose a challenge to $\Lambda$CDM, as the highest reported slopes are more consistent with MOND’s prediction of $M \propto V^4$ \citep{lelli16b}, and the very small intrinsic scatter of the observed BTFR is perhaps smaller than expected given variations in halo spin parameter, concentration, and baryon mass fraction \citep{mcgaugh12}. However, the cosmological significance of any best-fit BTFR is muddled by the relationship between observed and fundamental properties, observational uncertainties, and sample selection effects, making it difficult to fully assess the significance of these interpretations \citep{desmond17} and whether or not the scatter is consistent with expectations from $\Lambda$CDM \citep{desmond12}.

Aside from its cosmological significance, the observed low intrinsic scatter of the BTFR is valuable for measuring low-uncertainty redshift-independent distances to galaxies. Recent works have used this property of the TFR and BTFR to make local constraints on $H_0$ (\citealt{schombert20}, \citealt{kourkchi22}), and to measure large scale structure via peculiar velocities (e.g., \citealt{springob16}). The TFR has been used extensively to measure peculiar velocities and large-scale structure in the local universe (e.g., COSMICFLOWS (\citealt{kourkchi20}), \citealt{springob07}). Despite this, the curvature of the TFR at low mass indicates the utility of the BTFR for determining distances to a larger fraction of the (preferentially low-mass) sources in the local universe.

One application of BTFR distance estimates and a key focus of this paper is to measure the mass density of dark matter in large scale structure. \citet{eisenstein97} use the redshift dispersion of galaxies along the Pisces-Perseus supercluster, fortuitously aligned in the plane of the sky, to estimate the mass to light ratio of the supercluster within a factor of 2. Today, simulations suggest that infall onto filamentary structure can be detected using a sample of $\sim$few thousand galaxies with typical BTFR-derived distance uncertainties of $\sim$0.2 dex \citep{odekon22}. Measuring distances to this accuracy for such a large sample requires the use of a carefully calibrated BTFR. 

Aside from the cosmological considerations above, the slope, normalization, and observed scatter of the BTFR vary with the use of different velocity definitions (e.g., \citealt{lelli19}, \citealt{verheijen01}), selected galaxy populations \citep{bradford16}, methods for line fitting \citep{weiner06},  mass-to-light ratios used \citep{ponomareva18, mcgaugh14}, the size of the galaxy sample used \citep{sorce16}, and more. As with the Malmquist biases observed to increase the slope of the traditional TFR \citep{strauss95}, flux-limited samples may be biased by sensitivity-based selection effects \citep{kourkchi22}. Though there is general agreement that the slope of the BTFR is somewhere between $\sim3-4$, the heterogenous samples, types, and treatments of data used to calibrate literature BTFRs are partially responsible for the plethora of observed slopes.

Given its rich, homogenous, digital dataset, the Arecibo Legacy Fast ALFA (Arecibo L-band Feed Array) Survey (ALFALFA; \citealt{giovanelli05, haynes18}) presents fertile ground for an inclusive BTFR calibrated using uniform survey data. Here we use the ALFALFA extragalactic catalog \citep{haynes18} to derive a template relation specifically for use estimating secondary distances. Over the course of this work, we compare this fiducial template relation to others calibrated using different subsamples and velocity and mass definitions to investigate the influence of these choices on the presented relation.

In section 2, we review the sources of data and fitting methods used to find linewidths and BTFRs. In section 3, we discuss the methods used to fit our template relation, including considerations for sample selection and line-fitting methods for determining a template relation. We present a template relation using a high-quality, representative subsample of sources in section 4, identifying common classes of fit-identified outliers, comparing distance measurements to those from flow-models, and reporting typical distance uncertainties. In section 5, we compare this relation to others in the literature, finding that these relations become more consistent when demographics, velocity width measurements, and stellar mass estimating methods are matched. Section 6 presents an initial application of our derived distances -- measuring infall onto Virgo. Finally, we present our conclusions, including recommended guidelines for applications of this relation, in section 7.

\section{Data Compilation and Spectral Profile Fitting}
\subsection{Neutral Hydrogen (HI) Data and Gas Mass}
 The ALFALFA survey is a drift-scan, sensitivity-limited survey using the seven-beam ALFA receiver on the Arecibo 305m telescope. This survey detected 31,000+ extragalactic sources \citep{haynes18}, with the goal of sampling the \HI mass function to $\simeq 100$ Mpc. The typical sensitivity of the ALFALFA survey ($5\sigma$ detection of 0.75 Jy km/s for $W_{50} = 200$ km/s) allows for the detection of low mass (log($M_{HI}/M_\odot) < 8$)) sources out to $\sim 35$ Mpc and higher mass sources to $\sim 260$ Mpc  (heliocentric velocity = 17912 km/s). When calibrating the template relation for this study, we use the distance estimates from the ALFALFA catalog, which are a combination of flow model distances and distances from cluster catalogs \citep{haynes18}. The sample used in this analysis is drawn from the ALFALFA extragalactic catalog with the goal of being representative of the same population. The \HI masses and velocity widths used in this analysis are calculated using the $\alpha.100$ catalog and ALFALFA spectra respectively. The $\alpha.100$ catalog also contains measures of profile velocity widths including $W_{50,P}$ (defined as the width of the profile at 50\% of the peak flux), which is calculated with some manual intervention. The new widths presented here are calculated fully automatically and thus can also be homogeneously calculated for samples where there is interest in applying this analysis.

Assuming optically thin emission, \HI mass is computed using ALFALFA catalog flux values and the following standard formula:
\begin{equation}
    M_{HI} = 2.356\times10^5 D_{HI}^2 S_{21} \mathrm{M}_\odot
\end{equation}
where $D_{HI}$ is the distance to the source taken from the ALFALFA catalog (as described above) in Mpc,  and $S_{21}$ is the integrated \HI line flux in Jy km/s. Previous works find that  the inclination-dependent self-absorption correction is likely only $\sim 10\%$ of the \HI mass in most ALFALFA galaxies, and reaches $\sim 35\%$ only in the most edge-on sources \citep{jones18}, so no correction for self-absorption is made in the fiducial relation. 
To account for other contributions to the gas mass, the fractions of metals and molecular gas are incorporated via the flat conversion $M_{gas} = 1.33M_{HI}$, which is commonly used in BTFR studies. 

\subsection{Stellar Mass}
The stellar masses used in the fiducial relation are based on Sloan Digital Sky Survey (SDSS;\citet{blanton17}) optical photometry and are tabulated using equation (7) of \citet{taylor11} and additional considerations discussed in \citet{durbala20}. These masses are recommended among those discussed in \citet{durbala20} based on their agreement with rigorous SED-based mass estimates and the availability of data for a large fraction of sources in the full ALFALFA catalog. Additionally, NIR masses tabulated using unWISE \citep{lang16} photometry are subject to shredding effects leading to underestimated masses in the most extended ALFALFA sources -- SDSS-based stellar masses are not subject to these effects.
section~\ref{sec:stellarmass} presents a BTFR using unWISE NIR photometry-based stellar masses, which may be used to extend this work beyond sources with available optical observations.

\begin{figure*}
    \centering
    \includegraphics[width=0.49\linewidth, trim = 65 65 65 65]{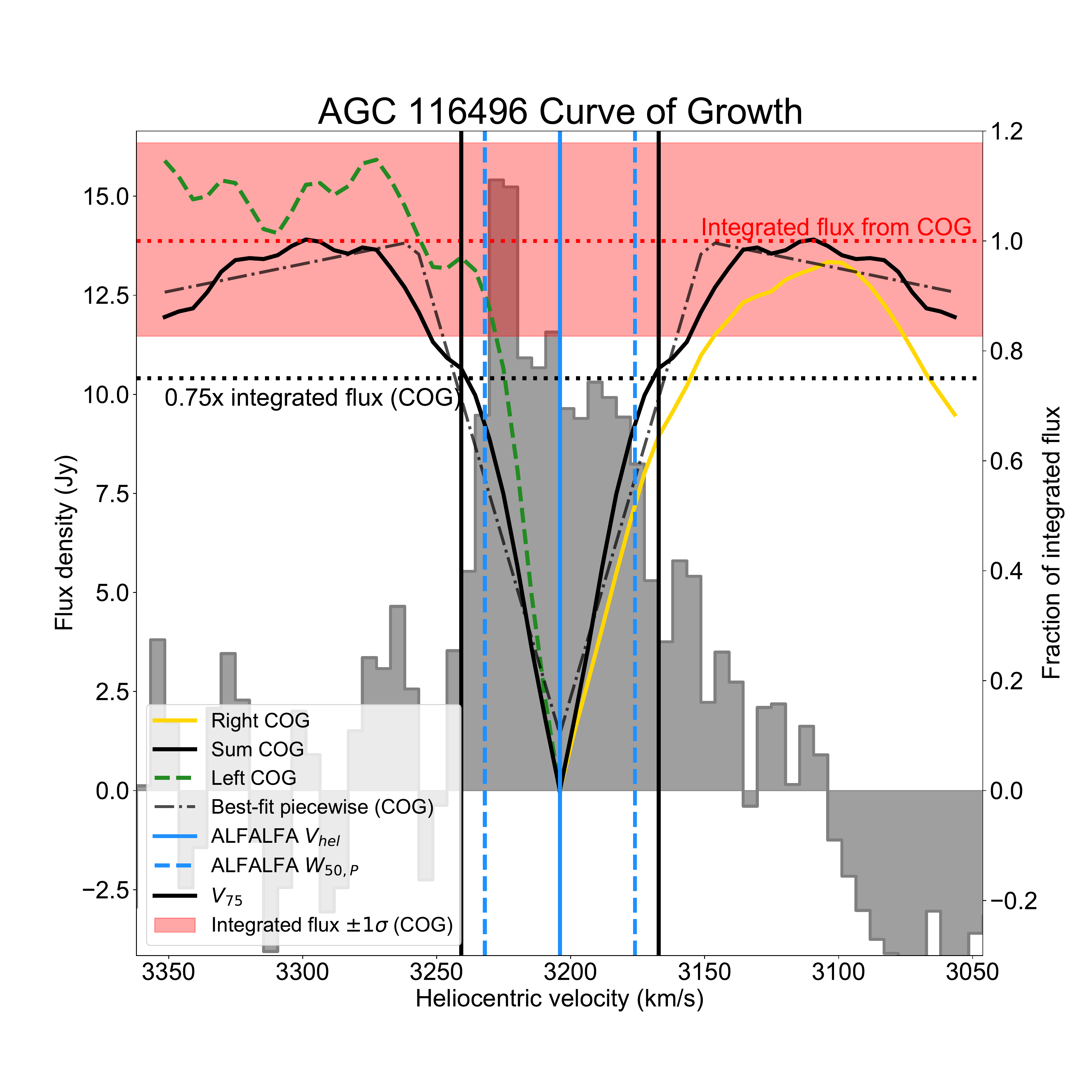}
    \includegraphics[width=0.49\linewidth, trim = 65 65 65 65]{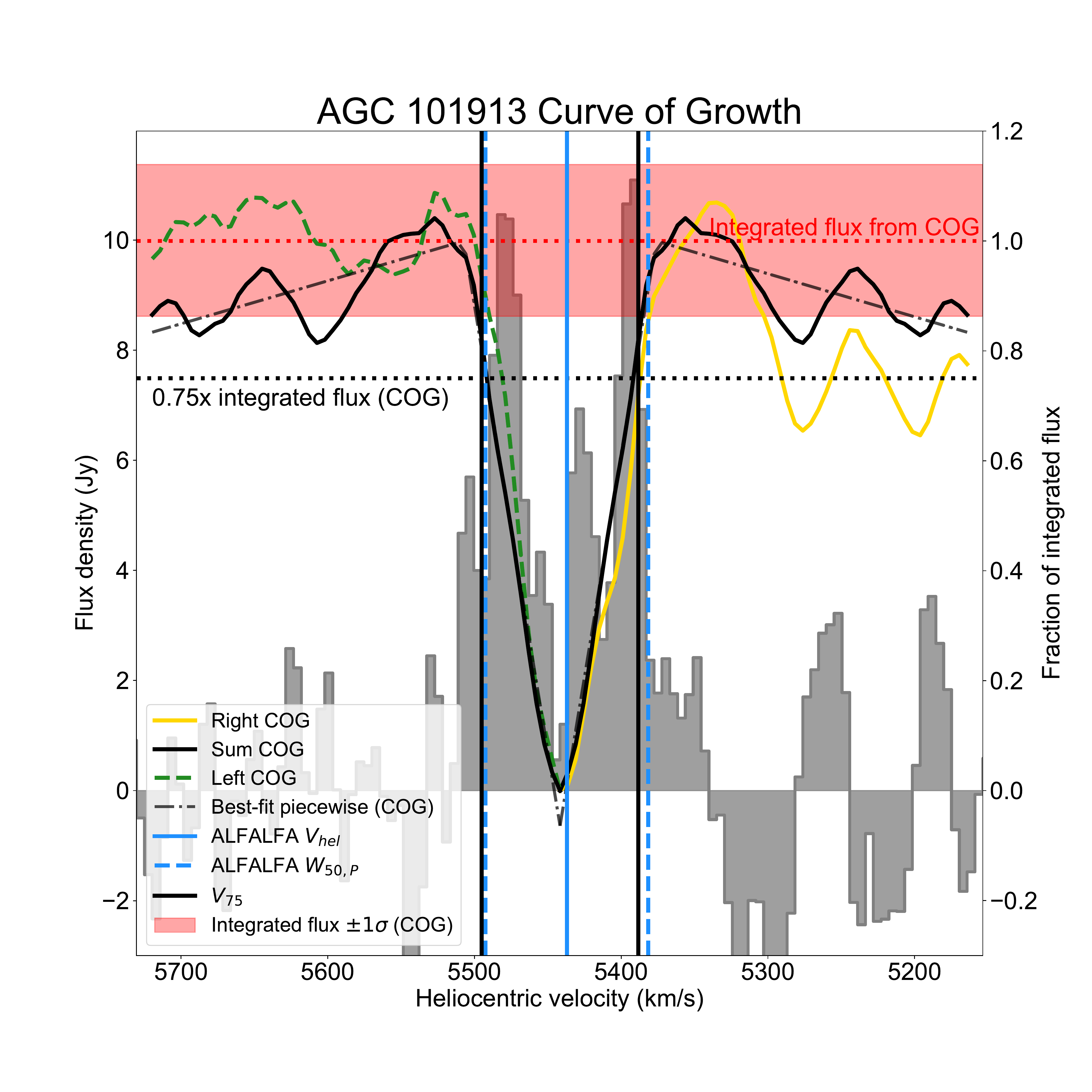}
    \caption{Output diagnostic plots for the automatic curve of growth width-fitting program described in section~\ref{sec:veldef} for three example galaxies, illustrating the performance of the program for sources with common shapes and typical S/N values. (Left) AGC 116496 is a single-horned, asymmetric profile with ALFALFA S/N $= 10.2$
    and (right) is AGC 101913, a double-horned profile with S/N $= 7.3$. Each panel includes the ALFALFA profile (grey) for each source. From the center channel of each profile, identified in step 1 of the procedure defined in sec.~\ref{sec:veldef}, the summed curve of growth (step 2) is plotted in black and is normalized by the integrated flux of the curve.  For illustrative purposes, these images also include the left and right curves of growth summed separately (green and yellow), with each of these curves normalized by their integrated flux in order to highlight the difference in shape of the curves. The integrated flux used to normalize the black curve of growth is $\sim2\times$ as large as the value used to normalize the left and right curves separately. The difference of the shape of these curves is most dramatic for asymmetric profiles (left) or low-S/N profiles, where using the summed COG improves the algorithm's ability to correctly identify the integrated flux in the profile. The red region indicates this integrated flux, indicated with the red dotted line, $\pm$ the propagated error. This integrated flux is determined using the best-fit piecewise function (e.g., step 3, with the piecewise fit indicated using the black dashed-dotted line in the figure), and the dotted black horizontal line indicates where the curve of growth crosses 75\% of this flux value (step 4). The solid, vertical black lines demonstrate where $2\times V_{75}$ appears on the profile. $W_{50,P}$ is also indicated for each profile to serve as a comparison -- as illustrated in Figure~\ref{fig:flag_v75s}, $V_{75} > W_{50,P}$ in the narrowest profile (left), and $W_{50,P}$ is larger in the profiles where the flux is more heavily weighted to the outer part of the profile -- e.g., the more flat-topped (center) and double-horned (right) profiles.}
    \label{fig:width_fit}
\end{figure*}

\subsection{Velocity Definition}\label{sec:veldef}
The doppler shift of the 21 cm line is a useful tracer of the motion of galaxies' baryons, and thus the rotational velocity at the extreme extent of galaxies. When resolved rotation curves are unavailable, a galaxy's rotational velocity can be approximated by measuring the width of the spectral line profile. To ensure uniformity, specific definitions of velocity width are used across the literature. Commonly used velocity width definitions measure the width across the global \HI line profile at some percentage of the peak flux (e.g., $W_{50,P}$), or similarly of the mean flux (e.g., $W_{50,M}$). Some velocity definitions are, in general, larger and in better agreement with the flat velocities at the extreme edge of a galaxy's rotation curve (e.g., $W_{20,P}$, \citep{lelli19}, $W_{50,M}$ \citep{courtois09}), others may be more robust at low signal-to-noise ($W_{50,P}$ \citep{springob05}). Additional velocity width measurements come from parametric fits to the full shape of a line profile (e.g., Busy Function, \citet{westmeier14}), or fitting linewidths using physical source modeling (Peng et al., {\em in prep}).

The velocity width used in this analysis is based on the curve-of-growth (COG) width-finding method proposed in \citet{yu20}. As discussed there, this algorithm is straightforward to implement automatically. Automatically measuring profile linewidths allows for efficient and uniform reduction of large datasets, which is our aim in this study. The program used here fits a piecewise function to a pair of lines -- the cost surface of this problem is generally simple, and relevant parameters (the pivot channel, the slope of the line on either side of the pivot) are straightforward to initialize with rigorous guesses\footnote{Such guesses are important as this is a nonlinear least squares problem, meaning the cost surface can be complicated and a simple optimization routine may find a local minimum which is not the global best fit}. Because of this, this method is computationally faster than a non-linear least squares fitting algorithm which include more parameters (e.g., the Busy Function \citep{westmeier14}; \citealt{stewart14}), and is currently more reliable than recent attempts using trained neural networks (e.g., \citealt{hallenbeck19}). Additionally, its dependence on the integrated flux of a line profile rather than its peaks and boundaries means its performance at low-S/N is significantly more reliable than attempts at automating $W_{50,P}$ and $W_{20,P}$. 

\begin{figure}
    \centering
    \includegraphics[width = \linewidth, trim = 75 75 75 75]{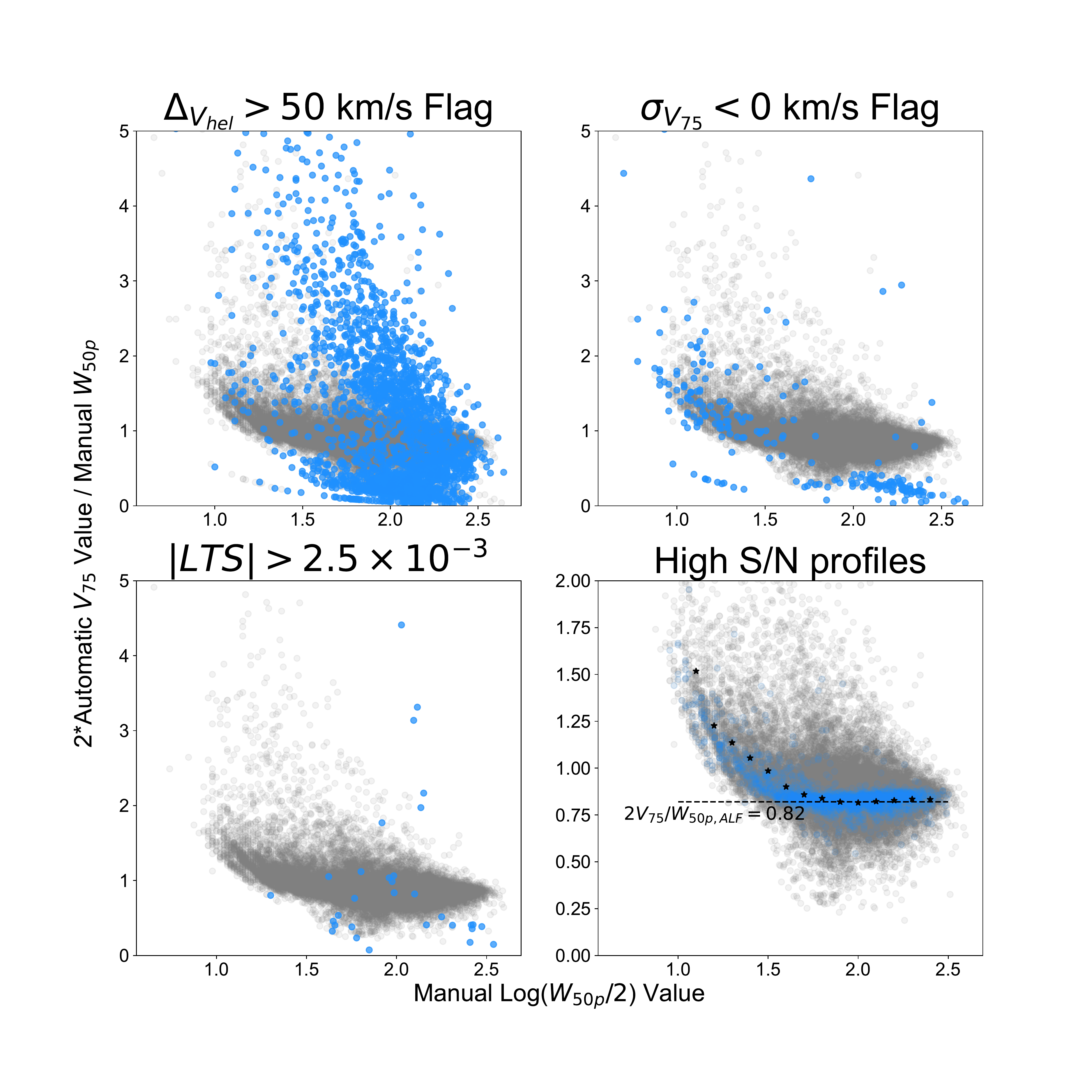}
    \caption{This figure demonstrates the removal of poor width measurements by the criteria discussed in the text. (Top left) Profiles highlighted in blue have $\Delta V_{hel} > 50$ km/s and are removed from the selection. (Top right) Profiles highlighted in blue have negative uncertainties and are removed from the selection. (Bottom left) Profiles highlighted in blue have curves of growth which do not appear to ``level out", as identified via steep long term slope magnitudes and are removed from the selection. (Bottom right) This shows the recommended profiles e.g., the profiles with the three other selection criteria applied, with profiles with ALFALFA S/N $> 30$ highlighted in blue in order to indicate that this relationship between $V_{75}/W_{50,P}$ is well defined, but scattered by noise and the success of both width-finding methods. The black stars are the mean $2V_{75}/W_{50,p}$ values per 0.1 bin in log($W_{50,p}$) space. The horizontal line shows a flat ratio $2V_{75}/W_{50,p} = 0.82$, which is in good agreement with the mean value of this ratio for sources with log($W_{50,p}) > 1.8$.}
    \label{fig:flag_v75s}
\end{figure}

The implementation of a curve of growth width-finding algorithm presented here is slightly different from the original presentation in \citet{yu20}.
Three examples of the fitting procedure used in this analysis are presented graphically in Figure~\ref{fig:width_fit} for galaxies with different profile shapes and typical S/N values, and a general sketch of the procedure is:
\begin{enumerate}
    \item Find the heliocentric velocity of a line profile using the hermite polynomial matched filtering described in \citet{saintonge07}. This determines the velocity channel where the two black curves of growth meet and are centered in the figure.
    \item Moving outward from the central channel, add the flux in subsequent channels to both the left and right in order to create a curve of growth - the integrated flux of the profile as a function of channel from profile center. The result of this integration is the black curve in Figure~\ref{fig:width_fit}. Errors for this curve are determined from the sum of the per-channel RMS. 
    \item Fit a piecewise function consisting of two straight lines to the resulting curve of growth, allowing for a coarse determination of the integrated flux of the profile -- we use the pivot of the piecewise as this integrated flux. This best-fit piecewise is plotted on each profile in Fig.~\ref{fig:width_fit} using a dashed-dotted line, and the pivot of the piecewise intersects the dotted red line, indicating this integrated flux. The red region in each figure is centered on the best-fit integrated flux value, with the region extending to $\pm1\times$ the uncertainty on that integrated flux value.
    \item To measure the COG velocity width, which is defined as the width of the profile at some percent of the integrated flux of the profile, the COG is renormalized, and the profile width is determined by the velocity width where the curve of growth summed from the data intersects the desired percentage of the total flux. In Figure~\ref{fig:width_fit}, this occurs when the black curve of growth intersects the dotted horizontal black line at 0.75 of the integrated flux, with the profile width indicated using solid black vertical lines on the figure.
\end{enumerate}
In theory, the summing of the curve of growth and subsequent piecewise fitting could be done separately for each half of the profile in order to more flexibly measure the widths of very asymmetric profiles. Testing this procedure on a subsample of sources finds that the widths are more reliable (i.e., follow the same general trends between $V_{75}$ vs. $W_{50,P}$ confirmed with a high S/N sample and shown in Figure~\ref{fig:flag_v75s}) when both sides are summed simultaneously -- in the typical S/N range of this sample, the increased S/N of the summed COG is more important than carefully treating the profile shape.

Occasionally, this width-fitting program outputs widths that are clearly discrepant from ALFALFA width measures (where the fit $V_{75}$ is more than twice, or less than half, the rotation velocity measured using $W_{50,P}/2$). The most robust indicator that the width-fitting has failed is if the output $V_{helio}$ measurement greatly differs from that of the ALFALFA catalog -- we use $\Delta {V_{hel}} = V_{helio, ALFALFA} - V_{helio,WF}$ where $V_{helio, ALFALFA}$ is the catalog measurement and $V_{helio, WF}$ is the measurement made by the width-fitting program. When $\Delta V_{hel}$ is large, the matched filtering step finds some other noise peak, or uses a single peak of a strongly double-horned profile as its center. We find that taking a cutoff of $|\Delta {V_{hel}}| < 50 ~km/s$ removes the majority of such sources from the sample, as illustrated in the first panel of Figure~\ref{fig:flag_v75s}. In this panel, the blue points are the sources selected with $|\Delta {V_{hel}}| > 50 ~km/s$, superimposed on the remaining sources which have  $|\Delta {V_{hel}}| \leq 50 ~km/s$ criteria (grey). As demonstrated in the following panels, two additional cuts further reduce the number of outliers created by the width-fitting program - (1) negative uncertainties, indicative of an issue in normalizing the curve of growth and (2) a long-term slope of the curve of growth above an empirically-determined value, which indicates that the piecewise curve fit did not find the flat part of the curve of growth successfully. These three exclusion criteria remove many of the most clearly incorrect $V_{75}$ values for the $\alpha.100$ sample, minimizing the number of good fits that are removed as we control for the performance of the width-fitting program. In total, these criteria flag 2539/31502 ($\sim 8\%$) ALFALFA sources, reducing the number of sources with $V_{75}/W_{50,P}$ ratios vastly different from the average range of $0.5 < 2V_{75}/W_{50,P} < 2$ (bottom right panel, Figure~\ref{fig:flag_v75s}).

In order to have confidence that the reported uncertainties represent the width-finding program's success in finding the true width of a line profile, we test the program on a sample of real profiles with S/N $>$ 300 by degrading them to lower S/N values. Comparing the high-S/N reliable values of the profile widths with those measured in the noisy profiles, $\gtrsim 68\%$ of these noisy profile widths agree with the “base truth” value within $1\sigma$ for S/N $>$ 7. The number within 2 and 3 $\sigma$ is nearly consistent with gaussian distributions above this S/N threshold as well, although for some classes of profiles there is an outlier fraction of $\sim$few percent. This suggests that, although the reported uncertainties are reasonably good at describing the difference between true and noisy profile widths, $\sim2\%$ of widths for profiles with S/N $>$ 9 may be inconsistent with expectations because of poor performance of the width-finding program. The outlier fraction increases in the lower-S/N regime of the sample, which have  4.5, 6, and 10\% sources with $\Delta V > 3\sigma$ for S/Ns 9, 7, and 5, respectively. Given the distribution of profile S/N values for the full ALFALFA sample and the sample used to fit the template relation (5.8, 8.4, 17 and 6.8, 10.2, 21 for the 16th, 50th, 84th percentiles, respectively), we expect $\lesssim6.6$\% of sources to have widths significantly distorted by noise in an ALFALFA BTFR and $\lesssim4$\% of the sample used to fit the template relation ($\alpha.R$) to have distorted widths attributable to the performance of the width-finding program.

An additional factor that makes it difficult to reliably measure the width of low-S/N profiles is variation in the baseline of the spectrum. To determine how the width fitting program is affected by baseline artefacts, we simulate noise and simple baseline variations over noiseless profiles simulated from input model rotation curves using the schema discussed in \citet{obreschkow09}. Again, we find similar consistency between the true widths and degraded widths as in the ALFALFA data experiment, as the piecewise fit in the width-fitting program is appropriately insensitive to artefacts beyond the bulk of the line (as can also be seen in Figure~\ref{fig:width_fit}). In the case of higher-order baselines (sloped, curved baselines), the method for calculating the uncertainty of the pivot flux takes these variations into account in determining appropriate uncertainties for measured widths.

Though $V_{75}$ is a smaller width than $W_{50,P}$, the two have a relatively constant proportionality for most double-horned profiles, with $V_{75}/W_{50,P}$ increasing at low widths where profiles are more Gaussian (see Figure~\ref{fig:flag_v75s}). This suggests that $V_{75}$ weights the shape of the line profile differently in determining the line profile width than $W_{50,P}$, as one would expect given that one looks at the integrated flux of the profile and the other looks at the extrema. Though in general smaller velocity measurements represent less extreme velocities in a galaxy's rotation curve, a straightforward flat multiplier correction ($V_{75} = 0.41 W_{50,p}$; see Figure~\ref{fig:flag_v75s}) could be applied to recover $W_{50,P}$ from $V_{75}$ across the majority of the ALFALFA mass range (for sources with log$(W_{50,P})> 1.8$, true for $\sim$26000/31000 ALFALFA galaxies). In fact, the increase in $V_{75}/W_{50,P}$ at low mass suggests that $V_{75}$ may in fact be more effective at tracing the extreme velocities at the edges of dwarf galaxies' slow-rising rotation curves (e.g., \citealt{mcquinn22}). Given the difficulty tying the width of a profile to the internal dynamics of a source, the differences between this velocity width and others does not preclude the use of $V_{75}$ in BTFR studies, rather it provides a new avenue for automatically fitting widths for use in large datasets.

\begin{figure}
    \centering
    \includegraphics[width=\linewidth, trim = 75 75 75 75]{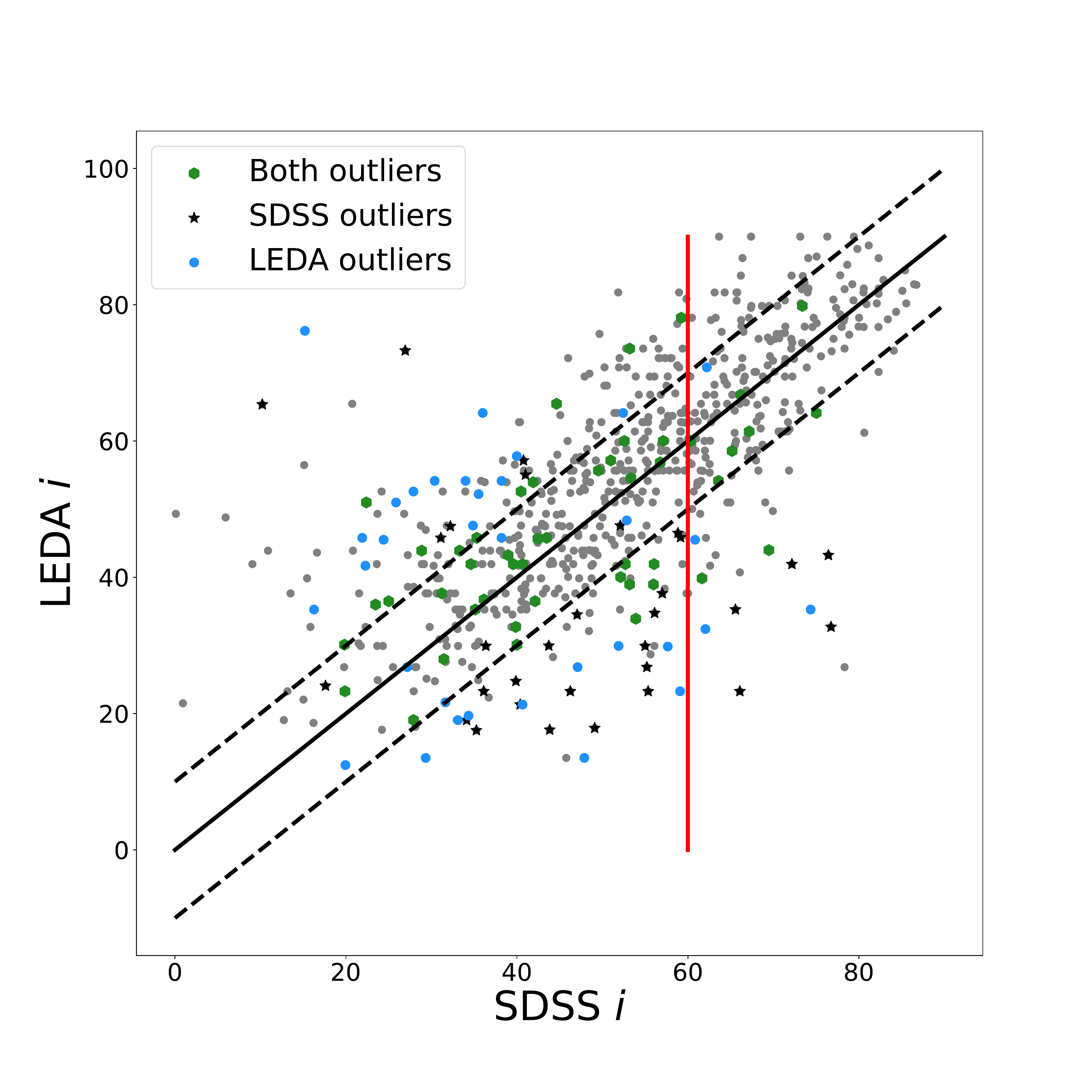}
    \caption{Comparison of inclinations (in degrees) in the HYPERLEDA-ALFALFA crossmatch sample. $x$ axis inclination values are derived using SDSS axial ratios and a constant assumed intrinsic axial ratio, $y$ inclination values are based on HYPERLEDA and a morphology-dependent correction factor. The solid black line is a 1:1 comparison, where the dashed black lines are offset by 10 degrees in each direction to highlight the typical scatter in the sample. Though we correct our velocities for inclination using sin($i$), we plot $i$ here in order to demonstrate the typical spread in inclination estimates, which motivates the addition of the $10^\circ$ systematic uncertainty term. Grey dots are all sources not identified as outliers in BTFRs fit to this crossmatch sample. Green dots are all sources which are identified as outliers in both, and black and blue dots are sources identified as outliers when the HYPERLEDA and SDSS inclination corrections have been applied, respectively. As expected, sources with less consistent inclination measures between the two samples are more likely to be outliers in one sample than the other. Making a strict cut at SDSS $i = 60^{\circ}$ reduces the occurence of these outliers due to misatributed inclinations.}
    \label{fig:inc_comparison}
\end{figure}

\subsection{Inclination}\label{sec:incls}

The observed doppler-broadened line width measures only the radial component of the galaxy's rotation and must be corrected for disk inclination. We estimate the inclination of these sources using their observed axial ratios from SDSS photometry and the relation 
\begin{equation}
    \sin(i) = \sqrt{\frac{1-\frac{b}{a}^2}{1 - q_0^2}}
\end{equation} where $\frac{b}{a}$ is the observed axial ratio of the source and $q_0$ is the intrinsic ratio \citep{hubble26}. These inclinations allow us to correct the sources' rotational velocities for projection effects: 
\begin{equation}
 V_{rot,75} = \frac{V_{75}}{\sin(i)(1+z)}
 \end{equation} We use $q_0 = 0.15$ -- most studies use values between $q_0\sim0.1-0.2$ for blue spirals, which we expect to dominate the ALFALFA sample (e.g., \citealt{karachentsev17} figure 2, \citealt{yuan04}). To investigate the reliability of these inclination estimates, we compare them with estimates tabulated in the HYPERLEDA catalog\footnote{http://leda.univ-lyon1.fr/}, which derives morphology-dependent inclination values using
\begin{equation}
    \sin(i) = \sqrt{\frac{1 - 10^{-2\log(r_{25})}}{1 - 10^{-2\log(r_{0})}}}
\end{equation} where log($r_0$) varies with morphological Hubble type and log($r_{25}$) is the B-band 25 mag/arcsec$^2$ isophotal axial ratio \citep{paturel03}. We plot these inclination estimates against one another in Figure~\ref{fig:inc_comparison} and find general agreement, consistent with \citet{bradford16}'s finding that a morphology-dependent inclination does not appear to significantly affect the best-fit BTFR. 

As illustrated by the dashed lines around the 1:1 ratio in Figure~\ref{fig:inc_comparison}, there is an average spread of $\sim$10 degrees difference in the inclinations measured via these different methods for the sample of crossmatched HYPERLEDA-ALFALFA sources. This is incorporated as a systematic term in the uncertainty of the inclination correction. The statistical term of this uncertainty comes from the propagation of the axial ratio uncertainty in the photometry of each source, and is in most cases much smaller than the systematic term. 

As can be seen in Figure~\ref{fig:inc_comparison}, there are a number of sources with inclinations $>10^\circ$ disparate between measurement methods; the axial ratios for these sources are different between these datasets and thus have different inclinations and inclination-corrected velocity measurements. We find consistent relations fit using SDSS and HYPERLEDA inclinations when using an outlier-downweighting fitting method. Sources with $\delta i > 10^{\circ}$ between estimation methods are often outliers from only one of the two BTFRs, which we take as evidence that these inclinations are not reliable. Of the 31/842 sources which are only outliers in the SDSS sample, 28/31 sources have $\delta i > 10^\circ$. Similarly, 31 of the 35 outliers from the HYPERLEDA relation have $\delta i > 10^\circ$. Corrections based on an over/underestimated inclination will move points away from the central relation, making them more likely to be outliers from the BTFR. When we find outliers where the inclination is overestimated, this overestimated inclination will lead to smaller sin($i$) uncertainties, causing these unreliable inclinations to have disproportionately large influence when using a simple fitting likelihood that does not account for an outlier fraction as described in section~\ref{sec:fitting}. These sources are dealt with through the downweighting of outliers using a Gaussian mixture model and by imposing an inclination restriction designed to reduce the fraction of sources with unreliable SDSS inclinations in the HYPERLEDA-ALFALFA crossmatch sample ($i > 60^\circ$, red vertical line in Figure~\ref{fig:inc_comparison}). In the sample with i $> 60^{\circ}$, 5 / 353 sources are identified as SDSS-only outliers, and these sources all have HYPERLEDA $i <<$ SDSS $i$. We note sources with understimated inclinations from SDSS axial ratios have previously been noted as outliers in other studies of the BTFR \citep{catinella12}. 

Note that the fiducial relation does not include a turbulence correction for the rotational velocities as is occasionally done in BTFR studies, although we do discuss a BTFR with such a correction applied in section~\ref{sec:fit_veldef}. As this is the first attempt to use and interpret a $V_{75}$-based BTFR, applying a uniform correction to a velocity measure likely to be more sensitive to $V/\sigma$ than other width measures (see, e.g., the different relation between shape and width demonstrated by the $V_{75}/W_{50,P}$ relation) adds unnecessary complication for our specific application of these velocities.

\section{Fitting the Template BTFR}
\subsection{Fitting Algorithms}\label{sec:fitting}
We fit the relations discussed here using two likelihood functions. The first is a simple linear model in log-log space much like the ones used in \citet{lelli19} and \citet{papastergis16}. As these observations have significant uncertainties in both the x and y variables, using straightforward linear regression is known to introduce a bias to the slope of a fit relation \citep{willick95}. Generative likelihood models rigorously handle this while maintaining model simplicity. This model is an intrinsic gaussian distribution perpendicular to a best-fit linear relation in log-log space. This model assumes that the uncertainties, here approximated as gaussian, describe the probability of an observation given a true underlying point within this distribution (see appendix A for more details.) The other model used to produce the presented fiducial relation, which we call the gaussian mixture model, includes an additional ``outlier" component parallel to this best-fit relation. As discussed in \citet{hogg10}, this class of model can be used to reduce the influence of outlier sources on a population's characteristics. These sources are difficult to exclude from the sample using the parametric cuts described in section~\ref{sec:exclusions}. They may include sources with underestimated uncertainties, misidentified optical counterparts in the $\alpha.100$ sample, or sources with intrinsic properties which make them distinctly different from the regular rotators most adherent to the BTFR (e.g., blended sources, mergers, tidally disturbed galaxies). 

\subsection{Restrictions to the ALFALFA sample}\label{sec:exclusions}
Restricted samples of high quality, targeted observations are important for use of the BTFR in cosmological analyses, as they reduce observational uncertainties and probe the fundamental relation while minimizing competing effects. As discussed in \citet{bradford16}, the properties of a sample used to define a template are important for determining where that template is applicable. If restrictive cuts are not appropriately accounted for in the fitting process, they can limit the applicability of, and in some cases, can bias the resulting relation (see demonstrations of this in section~\ref{sec:comparison}). 

\begin{table}[]
\centering
\begin{tabular}{lc}
\hline
Sample                                 & Number of Sources \\ \hline
$\alpha.100$                           & 31502                 \\
(1) Recommended by width fitter        & 28963             \\
(2) SDSS stellar masses available  & 28267             \\
(3) Far from overdensities              & 17772             \\
(4) $i > 60^{\circ}$                          & 7455              \\
(5) Width uncertainties $< 20$\%              & 8821              \\
(6) Not asymmetric profiles           & 11630             \\
(1-6) $\alpha.R$                     & 4525             
\end{tabular}
\caption{Table of the number of sources which fit each of the exclusion criteria described in section~\ref{sec:exclusions}.}
\end{table}

To define the template sample here, we take a different approach -- we make minimal cuts to preserve the properties of the underlying $\alpha.100$ sample of sources. The sample $\alpha.R$ (R for ``restricted") which defines the fiducial template relation includes some cuts which make it more selective than the full ALFALFA catalog. However, these cuts are agnostic to the demographics of the galaxy population (e.g., gas fraction, stellar mass, morphology), and are instead motivated by a desire to reduce the number of outliers caused by improper fitting, low-quality data, or sources with disturbed morphologies/unlikely to be regularly rotating. We also run fits which relax the $\alpha.R$-defining restrictions, showing that broadly the full $\alpha.100$ sample is consistent with this template (see section~\ref{sec:consistent}). The criteria used to define the $\alpha.R$ sample are:
\begin{enumerate}
    \item {\em Sources recommended for use by the output of the width-fitting program.} As described in section~\ref{sec:veldef}, we determine a number of criteria that prune the most severe failures of the width-finding program.
    \item {\em Sources that have stellar masses derived from SDSS photometry in the \citet{durbala20} catalog.}
    \item {\em Sources far from large overdensities.} One intended application of this relation is to measure signatures of infall onto the Pisces-Perseus supercluster (PPS). We want to fit a template that is unlikely to contain infall signatures, necessitating the removal of certain regions of the sky from the sample. To this end, we also exclude Virgo, the other large overdensity in the ALFALFA survey range. Within the ALFALFA footprint, we exclude all sources that are at $22^\mathrm{h} < $ RA $< 3^\mathrm{h}$ to exclude PPS and all sources with RAs $12^\mathrm{h}08^\mathrm{m}-12^\mathrm{h}56^\mathrm{m}$, decs $2^\circ-18^\circ$, and velocities 350-3000 km/s, to exclude Virgo from the sample. 
\item {\em Sources with inclinations $i > 60^{\circ}$.} As discussed in section~\ref{sec:incls}, we use an inclination cut of SDSS $i > 60$ degrees as shown in Figure~\ref{fig:inc_comparison} in order to reduce the number of sources with unreliable SDSS inclinations. This does not remove all of the outliers with SDSS $i >$ HYPERLEDA $i$ but instead minimizes the fraction of sources with misidentified inclinations while maximizing the size of the remaining, non-restricted sample.
\item {\em Sources with uncertainties $< 20\%$ profile widths.} This restriction is essentially a width measurement S/N threshold. We test a number of uncertainty thresholds and find, as expected, that the slope of the relation increases as we impose a stricter uncertainty threshold and the noisy scatter of the distribution of widths is reduced. Above this imposed 20\% uncertainty threshold, the slope of the fit BTFR stays relatively constant. This threshold therefore appropriately limits the additional scatter and uncertainty caused by low-S/N widths and is the least restrictive cut we can make to this effect. 
\item {\em Sources which do not appear highly asymmetric to the width-fitting program.} Sources may appear highly asymmetric for a number of reasons. Firstly, this selects for many of the remaining failures of the width-fitting program -- if the program finds an inaccurate $V_{helio}$, it will center the curve on a noise spike or on one horn of a double-horned profile,  causing different ``widths" measurements for the left and right sides of the profiles when measured separately. Second, interacting sources may also have highly asymmetric line profiles \citep{espada11, bok19}. As these sources are disturbed, they are more likely to be offset from the BTFR defined by regularly rotating sources. Confusion (e.g., \citealt{jones15}) may also contaminate the sample -- this emission may appear lopsided if the two sources in the beam are at different distances from the \HI centroid.
\end{enumerate}

The imposition of all six of these criteria produce a reliable sample of $\sim 4500$ sources broadly representative of the $\alpha.100$ population, which we use to define a template relation that can be used as a distance indicator with minimal uncertainty and contamination.

\section{Results}
\subsection{Presentation of Relation}
The fiducial ``inclusive ALFALFA relation" presented here is based on the $\alpha.R$ sample discussed in section~\ref{sec:exclusions}. This relation is the bulk component of a Gaussian mixture model. As described in the next section, the majority of sources identified as outliers are not disky, isolated, ``typical" sources well-described by their catalog parameter values. The Gaussian mixture model downweights these outliers, prioritizing the majority of the sample which are more likely to be the well-behaved, rotating disks most adherent to the BTFR.

\begin{figure}
    \centering
    \includegraphics[width = \linewidth, trim = 75 75 75 75]{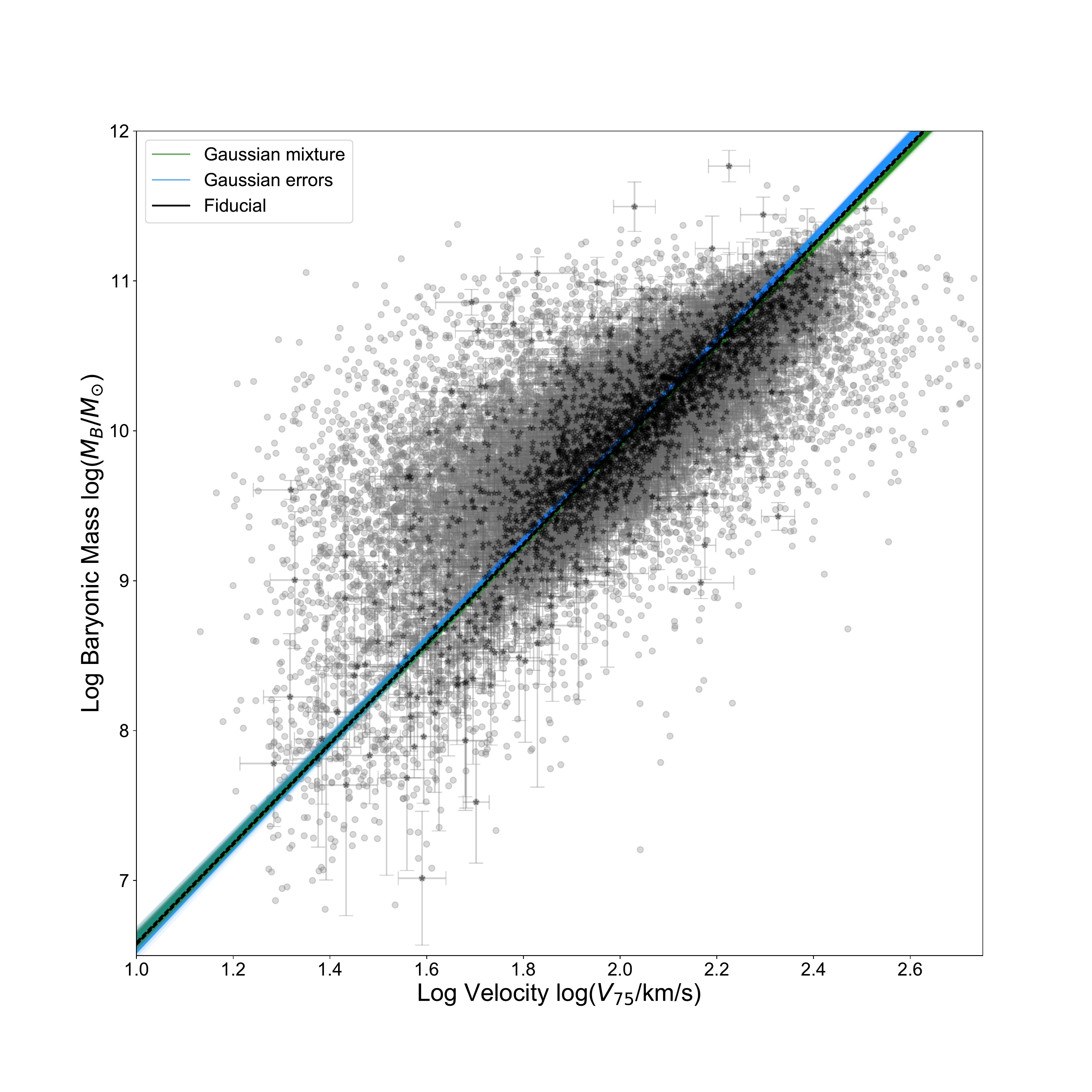}
    \caption{Template ALFALFA BTFR (equation~\ref{eqn:BTFR}) fit using the $\alpha.R$ sample (Black points) with the more general ALFALFA sample in grey beneath. Both models are overplotted, with the Gaussian mixture in green and the simple line in blue. These lines are plotted using random samples from the MCMC posterior output; the width of the lines demonstrates the range of probable best-fit models.}
    \label{fig:fiducial}
\end{figure}

The template $\alpha.R$ relation: 
\begin{equation}\label{eqn:BTFR}
   \log(M_{b}/M_\odot) = (3.30\pm0.06)*(\log(V_{rot,75}) - 2) + (9.9\pm0.014)  
\end{equation}
(where $V_{rot,75}$ is $V_{75}$ corrected for inclination and redshift broadening) has an intrinsic perpendicular scatter of $(0.025\pm0.004)$. This model fits for three additional parameters which describe the distribution of outliers -- this component is a gaussian distribution which runs parallel to the best-fit relation (equation~\ref{eqn:BTFR}) with a perpendicular offset ($-0.07\pm0.01, \sigma = 0.13\pm0.01$), and a weighting of ($0.2\pm0.03$) relative to the main component, which serves primarily to downweight the largest outliers. This relation is fit using 3000 galaxies randomly drawn from the 4525 galaxies in $\alpha.R$ to leave a significant sample of $\alpha.R$ sources available for validation. We note that this relation is stable within the uncertainties regardless of which random sample of 3000 sources is drawn. This template relation, along with a relation fit without a Gaussian outlier component, is presented in Figure~\ref{fig:fiducial} and the first row (Ia) of Table~\ref{table:btfr_fits}, which summarizes all of the fits run in this investigation. To the eye, the fiducial relation may appear steeper than the distribution of data; this is an expected result of rigorously fitting a relation to data with significant uncertainties in the independent variable -- fits by eye or using least-squares methods (which assume no $x$ uncertainties) will tend to fit relations that are inappropriately shallow.
 
 The uncertainties reported here include both the statistical uncertainties from the fitting procedure (reported in Table~\ref{table:btfr_fits}) and systematic uncertainties (0.05 in $m$ and 0.01 in $b$), estimated using the spread in best-fit relations when we relax the imposed restriction criteria in section~\ref{sec:exclusions}. Importantly, the slope and intercept of this template relation (equation~\ref{eqn:BTFR}) is consistent with these relations (see Figure~\ref{fig:consistency}). Though defining a restricted sample is useful for ensuring that the relation is fit with a high-confidence sample, the consistency with fits to less-restricted samples suggests that this $\alpha.R$ relation can be used to estimate distances to the broader $\alpha.100$ sample, including sources which do not fit all listed selection criteria. As discussed later in this section, the corresponding uncertainties and outlier fractions will correspondingly be larger in this case.

\subsection{Outlier classes from fiducial relation}\label{sec:outlier_classes}
This fiducial relation is drawn very generally from the full $\alpha.100$ catalog -- itself complicated by issues of varying coverage, the impact of RFI, confusion and uncertainties in the identification of optical counterparts -- meaning that there is a possibility of distorted data which may affect the relation and be difficult to remove parametrically. This motivates use of the mixture model to de facto downweight the most anomalous data. We investigate the 80 sources with the lowest probability of membership in the bulk relation model\footnote{See dfm.io/posts/mixture-models for more details.}, finding the following common classes:

\begin{figure}
    \centering
    \includegraphics[width = \linewidth, trim = 75 75 75 75]{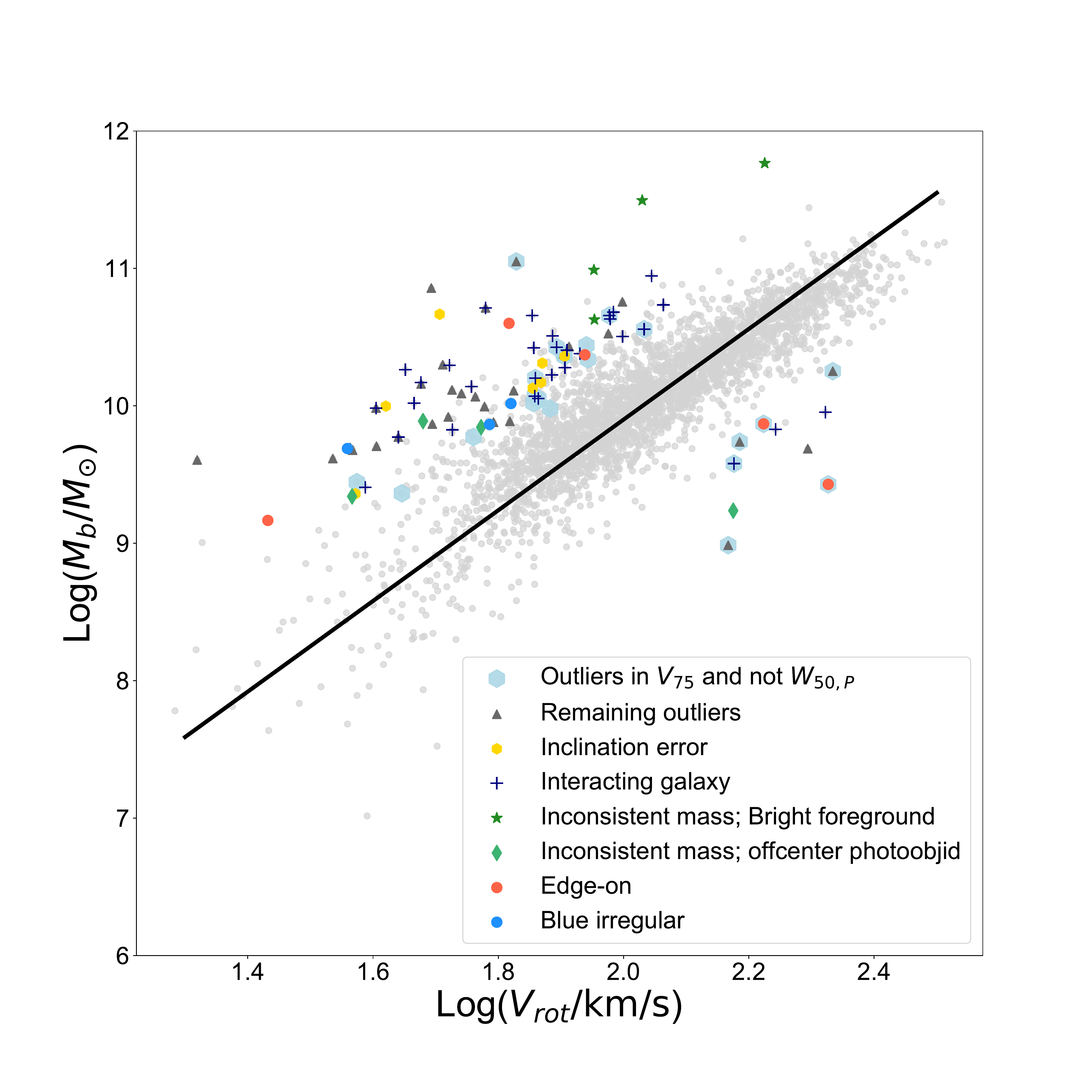}
    \caption{Outlier classes discussed in section~\ref{sec:outlier_classes}. The position of some of these outliers relative to the bulk relation is related to the class of outliers they fall into. For instance, the eight sources with overestimated inclinations (gold circles) sit to the left of the bulk relation because the tabulated, inclination-corrected velocity is too small. The four sources with bright foreground stars (green stars) sit above the bulk relation -- these sources appear brighter than they are intrinsically because of the starlight which has leaked into their images. Other classes of outliers have sources scattered on all sides of the bulk relation -- interacting sources (navy crosses) have a larger scatter around the bulk relation because of the complicated interplay of effects on both the baryonic mass and line profile. Sources with light blue hexagons behind the outlier symbol are not outliers in an $\alpha.R$ BTFR fit using $W_{50,P}$ instead of $V_{75}$.}
    \label{fig:outliers}
\end{figure}
\begin{enumerate}
\item {\em Sources with close companions.} Thirty sources ($\sim38\%$ of the 80 most extreme outliers) have nearby companions, identified in SDSS imaging by small sky ($\lesssim$ 6 arcmin, $\lesssim 120$ kpc at 70 Mpc) and velocity ($\sim$few 100s km/s) separations. Seven of these sources are deblended from nearby sources in the ALFALFA data reduction process. Nearby companions may impact sources' positions on the BTFR as confusion with gas-rich sources within the ALFALFA beam or distortion of \HI disks via interaction add significant non-rotational velocity to these source profile widths. Other studies find evidence that line profiles are affected by the presence of a close companion, including twice as many kinematic anomalies in pairs versus field galaxies \citep{kannappan04}, significantly more asymmetry in \HI line profiles for sources in close pairs vs. isolated sources \citep{bok19}, and more flat-topped or widened \HI profiles for sources in close pairs \citet{zuo22}. 
The outlying sources with close companions sit both above and below the bulk relation in Figure~\ref{fig:outliers}, demonstrating that sources with close companions add scatter to the bulk relation but do not preferentially displace these sources to larger velocity widths.

\item {\em Sources with misidentified inclinations} Eight ($10\%$) of the 80 extreme outliers can be confidently identified as nearly face-on in SDSS imaging. All eight sources have external evidence suggesting SDSS inclinations may be unreliable, including HYPERLEDA inclinations $>10^\circ$ different for 7/8 and detailed modeling in \citet{meert15} and Galaxy Zoo classification as `not edge-on' \citep{willett13} for the 8th. All eight of these sources sit to the left of the bulk relation in Figure~\ref{fig:outliers}, suggesting that their inclination-corrected velocity is underestimated, consistent with the interpretation that the 1/sin($i$) factor is too small for these sources. Though the inclination selection criteria in sec.~\ref{sec:exclusions} is designed to account for most outliers due to misattributed inclinations, a small fraction of overestimated SDSS-based inclinations in the above HYPERLEDA-ALFALFA comparison persists even with this selection criteria. The persistence of a (albeit small, $\sim$0.25\% of the total sample) fraction of misattributed inclination outliers is unsurprising.

\item {\em Sources with inconsistent stellar mass estimates.} Nine ($\sim11\%$) additional outliers have stellar masses that are likely to be unreliable. Four such sources have bright foreground stars in the same field as the galaxy in SDSS imaging. These sources all sit significantly above (large, positive $\Delta \log(M_b)$) the general relation, suggesting that their masses are overestimated, potentially appearing brighter than expected because of leaking flux from the foreground star. The remaining five sources without foreground stars have inconsistent $>5\times$ SDSS and unWISE stellar masses. These sources all have an SDSS photoobjid which is offset from the center of the source and instead centered on a compact region of brightness $<<$ the size of the galaxy, meaning that a misassigned photobjid may contribute to these sources' separation from the bulk relation. 

\item {\em Sources with inconsistent velocity measures.} Several sources may be outliers because of unreliable $V_{75}$ measurements. Twenty-one sources identified as outliers when fitting the $V_{75}$ relation are not identified as outliers when fitting the $W_{50,P}$ relation. Ten of these sources overlap with the above categories – 9/21 have close companions, 1/21 has an overestimated inclination. The remaining eleven sources represents a fraction of the $\alpha.R$ 3000-source sample (0.3\%) significantly smaller than the estimated percentage of $\lesssim4\%$ incorrect widths described in section~\ref{sec:veldef}. Even if the majority of these 21 sources are primarily outliers because their $V_{75}$ values are incorrect, this still indicates an outlier percentage $0.7\% << 4\%$. This closely-analyzed sample of 80 outliers represents only the most discrepant sources from the fiducial relation -- this underrepresentation of width-based outliers in this analysis may indicate that scatter due to inconsistent widths increases the scatter but is not significant enough to dramatically increase the number of outliers in the ALFALFA BTFR.
\end{enumerate}

The above four classes account for 58/80 of the sources identified as outliers by their low probability of membership in the bulk component of the gaussian mixture model. The remaining $\sim$quarter (22/80) of outliers are somewhat more complicated to interpret. Three are additional blue sources with irregular morphologies, three are the remaining highly inclined sources which may be affected by unaccounted for extinction / absorption corrections, but most (16/22) are too faint in SDSS imaging to interpret and are not particularly anomalous in other one- and two- parameter descriptions of the full ALFALFA dataset (e.g., gas fraction vs. mass, $V_{75}$ vs. $W_{50,P}$, S/N, etc.). Regardless, we are able to account for the majority of outliers from the fiducial relation, finding tractable reasons for their offsets.

\subsection{Residuals and expected accuracy of distance estimates}

\begin{figure}
    \centering
    \includegraphics[width = \columnwidth, trim = 75 75 75 75]{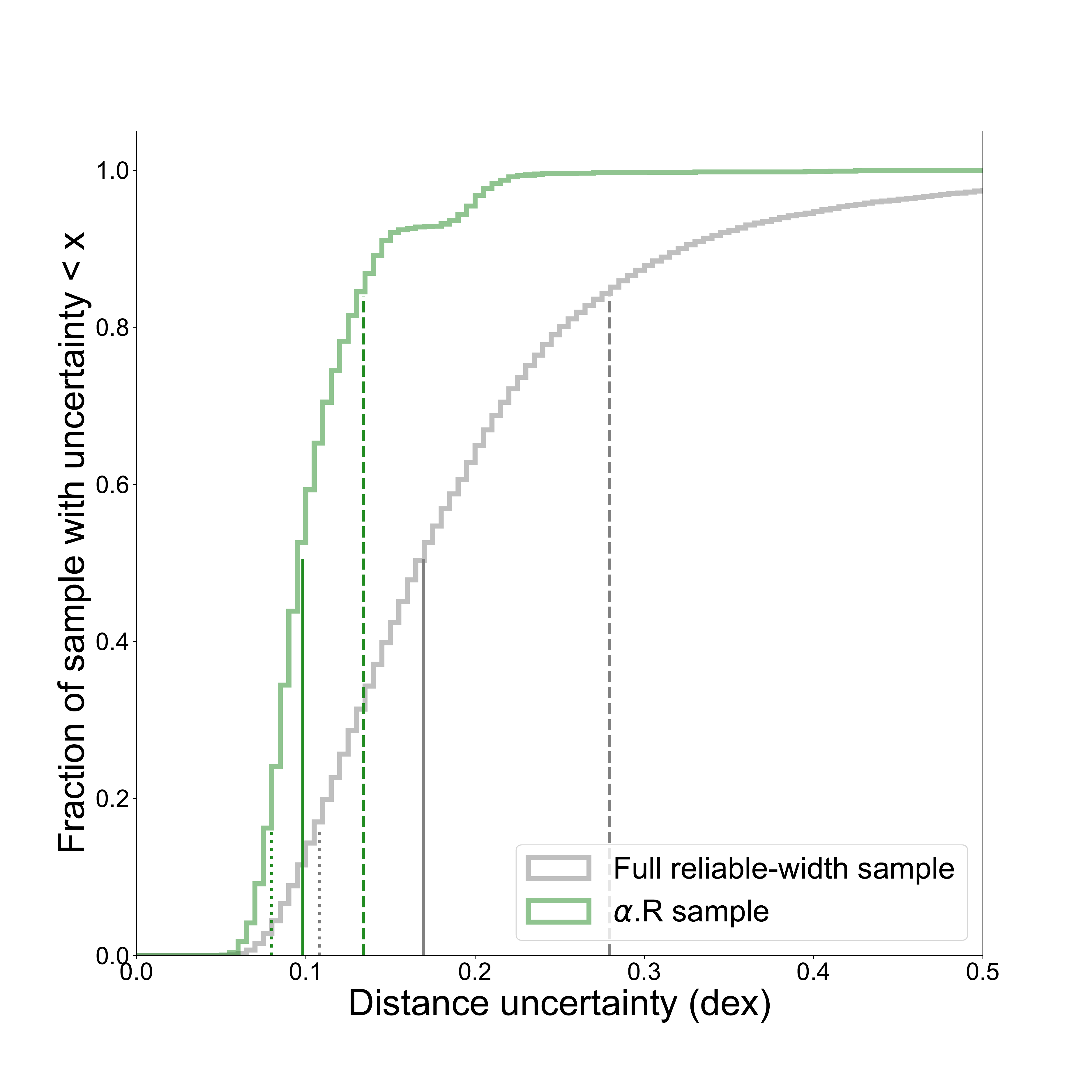}
    \caption{Reported distance uncertainties for the full ALFALFA sample (grey) and for the $\alpha.R$ subsample (green). The six vertical lines indicate where each curve crosses the 16th (dotted), 50th (solid), and 84th (dashed) percentiles, indicating the median uncertainty for each sample. The bump in the $\alpha.R$ curve at 0.2 dex is caused by the systematic uncertainty on ALFALFA HI flux measurements, which becomes the dominant uncertainty on log($M_b$) measurements at high-S/N -- in profiles more likely to be part of the $\alpha.R$ sample.}
    \label{fig:dist_uncs}
\end{figure}
\begin{figure}
    \centering
    \includegraphics[width = \columnwidth, trim = 75 75 75 75]{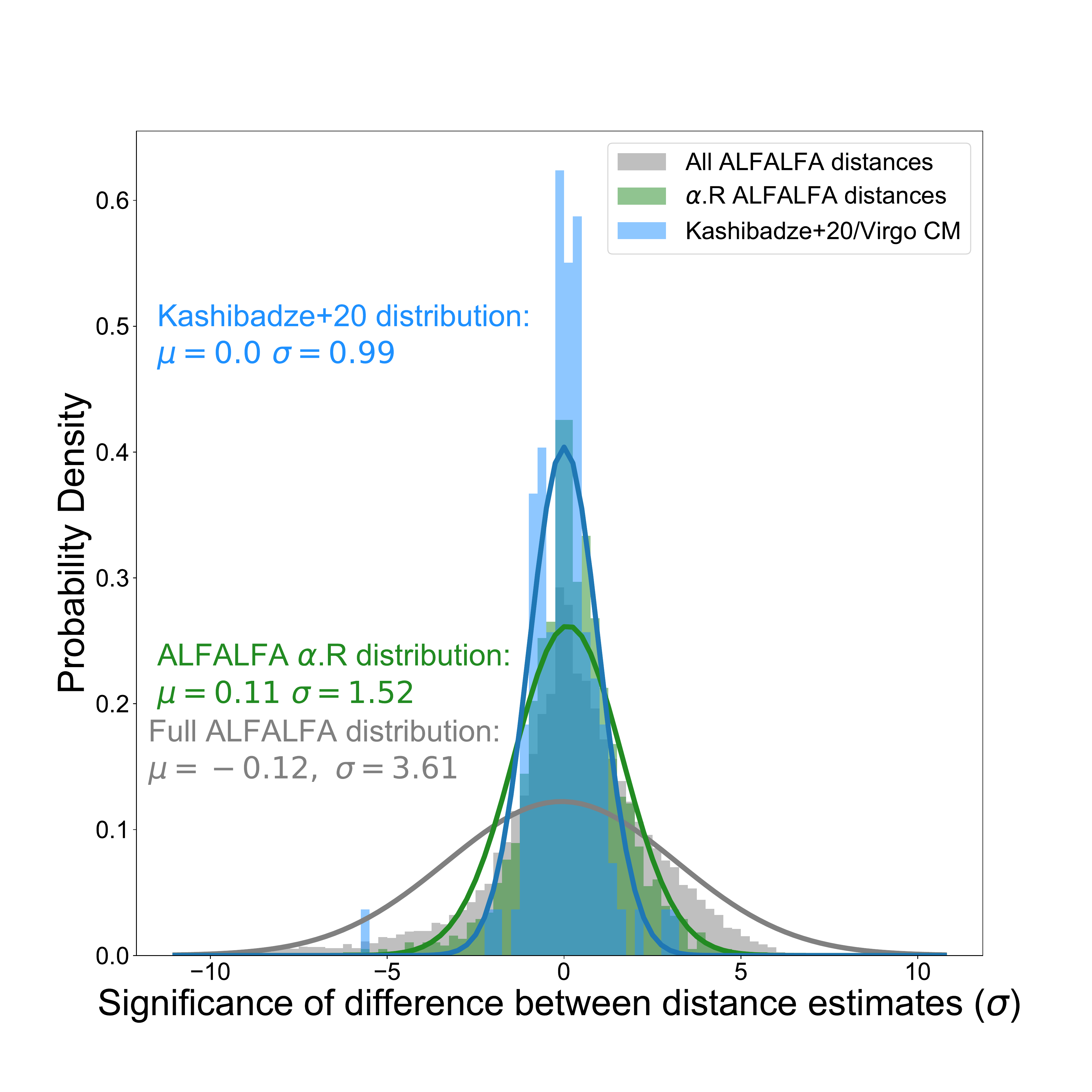}
    \caption{Comparison between BTFR distance estimates and estimates from other sources: the original ALFALFA catalog (grey; distances determined from a combination of flow model, cluster distances, and higher quality distance indicators; see \citet{haynes18} for details), ALFALFA distances for the $\alpha.R$ subsample (green), and a subsample of 109 galaxies in the vicinty of Virgo from \citet{kashibadze20} (blue). For the ALFALFA sample, the distribution is more heavy-tailed than the best-fit gaussian (solid lines in colors corresponding to the samples) owing in large part to the significant spread in BTFR-based distance estimates at large distances as well as the expected, nontrivial outlier fraction.}
    \label{fig:distance_comparisons}
\end{figure}
We use all ALFALFA extragalactic sources with recommended widths and \citet{durbala20} SDSS-based stellar masses to estimate the average uncertainty of a BTFR-derived distance measurement. For this full ALFALFA sample, propagation of error gives a median uncertainty on log(distance) estimates of 0.17 dex, with 16th and 84th percentile uncertainty estimates of 0.1, 0.28 dex. The cumulative distribution of uncertainties for BTFR-derived distances is shown as the grey curve in Figure~\ref{fig:dist_uncs}, with vertical bars at each of these reported percentiles. When we compare the predicted distance measurements to ALFALFA distances, which are calculated using flow models as described in \citet{haynes18}, we find that the distribution has a significantly larger standard deviation than would be expected if the departures were pure gaussian noise (e.g., $\mu = 0, \sigma = 1$). The grey distribution in Figure~\ref{fig:distance_comparisons} is the difference between the ALFALFA distances and the BTFR-based distances for the full ALFALFA sample -- the distribution is more sharply-peaked than the best-fit Gaussian, indicating the heavy tails of this distribution and the significant number of sources with discrepancies between these distance measurements -- $\sim$6\% of these distances are $> 3\sigma$ different. This is larger than expected if the differences are solely attributable to gaussian noise; however given the non-zero outlier fraction from Gaussian mixture modeling, the numerous causes of additional noise around the relation discussed above, and the corresponding increase in noise from looking at a less restricted sample we expect a larger-than-gaussian number of outliers. 

Using the more restricted $\alpha.R$ sample of sources, the median, 16th and 84th percentile uncertainties decreases to 0.092, 0.07 and 0.12, respectively (shown as the green curve in Figure~\ref{fig:distance_comparisons}, which reaches larger cumulative fractions at a given uncertainty than the grey curve), and the agreement with ALFALFA distances also increases as expected.  Overall, these distances are in much better agreement with the ALFALFA distances -- however, there are still about $\sim5\%$ sources with distance disagreements larger than $3\sigma$, and the heavy tails demonstrate the presence of outliers even in the restricted sample.

\begin{figure*}
    \centering
    \includegraphics[width = \linewidth, trim = 75 75 75 75]{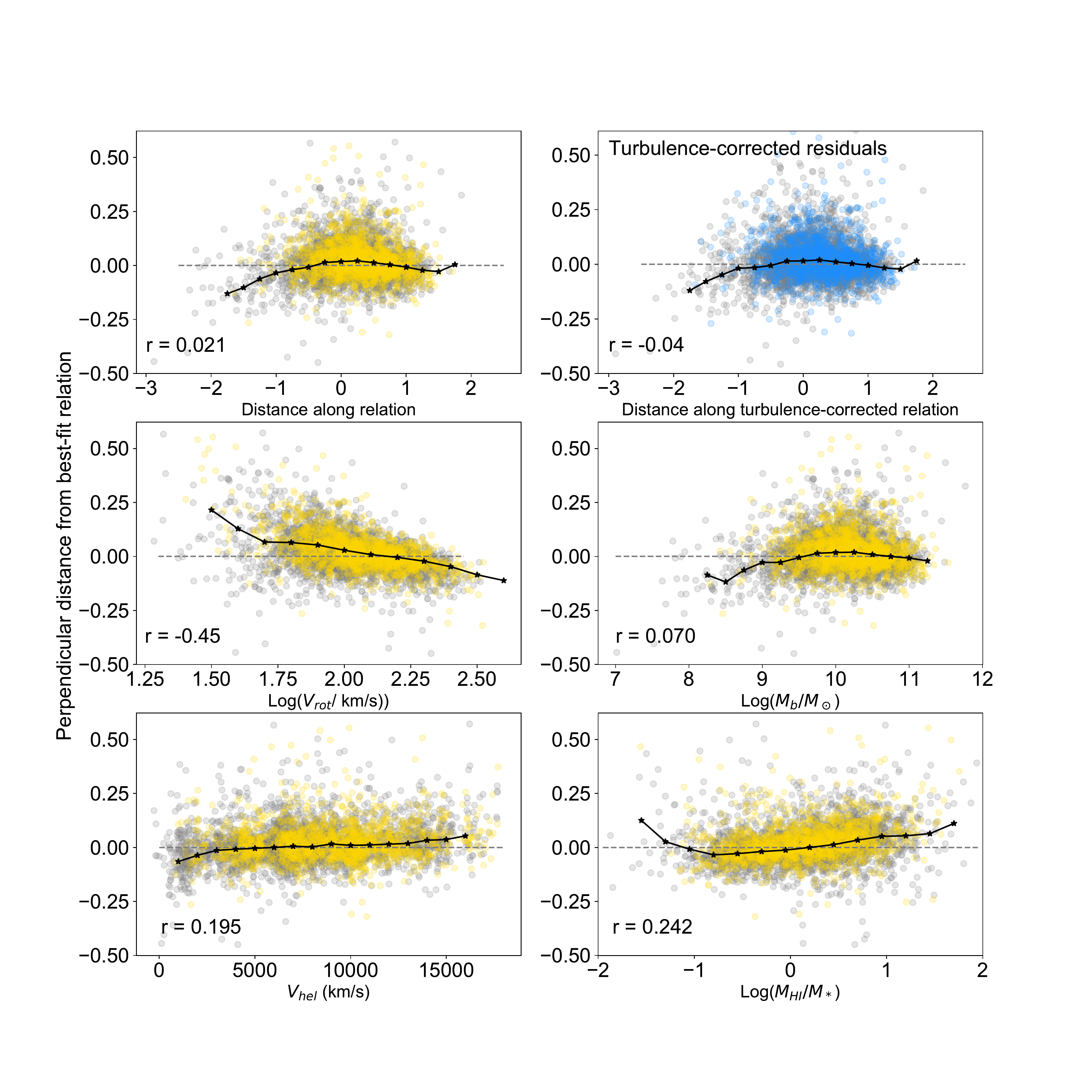}
    \caption{Perpendicular distance (positive values mean data points sit above best-fit relation) of template sample (3000 $\alpha.R$ galaxies) (grey) and the remaining $\alpha.R$ sources not used to fit the best-fit relation (gold). The top right figure demonstrates the marginal effects of including a turbulence correction on the relation (e.g., fit IIIb in Table~\ref{table:btfr_fits}) -- in this case, the points are based on the best-fit relation to $V_{rot,75} = V_{75,c}/((1+z)\mathrm{sin}(i))$, excising the 25 sources with negative $V_{rot,75}$ because the turbulence correction is overestimated in these sources. This relation shows a similar, slight downturn in the residuals at low velocity/mass, suggesting that this structure cannot be fully accounted for with a turbulence correciton. The remaining figures demonstrate the structure of these residuals with various source properties including distance along best-fit relation (top left), corrected rotational velocity (center left), baryonic mass (center right),  $V_{hel}$ (bottom left), and gas fraction (bottom right). The black lines on top represent binned median perpendicular distances, and the Pearson correlation coefficient for the full fiducial sample is located in the lower left corner of each figure.}
    \label{fig:res_trends}
\end{figure*}

Additionally, comparing the residuals from the template relation with various sample properties allows us to search for systematics which may be important when using this relation to measure source distances. We find that sources are generally distributed symmetrically about the relation as shown in the top left panel of Figure~\ref{fig:res_trends}, although there are hints of a fall-off at the lowest masses/velocities suggesting galaxies have larger velocities at a given mass. A similar trend appears in the residuals of the turbulence-corrected relation, suggesting that if the fractionally larger contribution of turbulence to velocity widths at low velocity causes this departure, the single turbulence correction as a function of velocity proposed in \citet{yu20} may not fully account for the variation of this correction across the galaxy population. The strongest correlation we find is between rotational velocity and perpendicular distance (left panel, second row of Figure~\ref{fig:res_trends}). This moderate correlation is likely attributable to the distribution of datapoints rather than to a slope mismatch for a number of reasons. First, a large number of sources with the largest positive distances do not sit at the lowest masses and rather at log($M_b) > 9.5$, suggesting that these sources have large perpendicular distances specifically because they have low rotational velocities for sources of their mass. We further test this by varying the slope of the relation and recalculating the correlation coefficients between the resiudals and velocity/mass, finding that as the magnitude of the $V_{rot,75}$ coefficient decreases, the magnitude of the log($M_b)$ coefficient increases -- if the correlation was due to a mismatch in the slope, both coefficients would decrease as we reached the correct value. In addition to this distribution-based correlation, we find additional, weak correlations between the residuals and the heliocentric velocity, a selection effect due to the flux-limited strategy of the ALFALFA survey \citep{kourkchi22} and with gas fraction, consistent with results from \citep{ponomareva18}. 

\section{Discussion}\label{sec:discussion}

\begin{table*}[]
\centering
\resizebox{\linewidth}{!}{
\begin{tabular}{ccccccccc}
\hline
ID   & Sample characteristics                                                                                   & Velocity Width & Simple Slope           & Simple Intercept & Simple Scatter            & Gauss Mix Slope        & Gauss Mix Intercept & Gauss Mix Scatter         \\ \hline
Ia   & Fiducial                                                                                                 & $V_{75}$       & $3.36\pm0.04$          & $9.94\pm0.01$    & $0.065\pm0.0025$          & $3.30\pm0.03$          & $9.9\pm0.01$        & $0.024\pm0.008$           \\
Ib   & Restrictions w/o 3                                                                                       & $V_{75}$       & $3.4\pm0.04$           & $9.96\pm0.01$    & $0.07\pm0.002$            & $3.34\pm0.03$          & $9.9\pm0.01$        & $0.026\pm0.004$           \\
Ic   & Restrictions w/o 4                                                                                       & $V_{75}$       & $3.43\pm0.04$          & $9.94\pm0.006$   & $0.072\pm0.002$           & $3.34\pm0.03$          & $9.89\pm0.01$       & $0.027\pm0.003$           \\
Id   & Restrictions w/o 5                                                                                       & $V_{75}$       & $3.32\pm0.03$          & $9.96\pm0.005$   & $0.0677\pm0.015$          & $3.28\pm0.03$          & $9.91\pm0.01$       & $0.026\pm0.005$           \\
Ie   & Restrictions w/o 6                                                                                       & $V_{75}$       & $3.33\pm0.03$          & $9.96\pm0.005$   & $0.071\pm0.002$           & $3.29\pm0.03$          & $9.9\pm0.01$        & $0.025\pm0.0045$          \\
If   & Forced same mass sampling                                                                                & $V_{75}$       & $3.37\pm0.035$         & $9.949\pm0.005$  & $0.064\pm0.001$           & $3.326\pm0.03$         & $9.91\pm0.01$       & $0.029\pm0.003$           \\ \hline
IIa  & \begin{tabular}[c]{@{}c@{}}Fiducial, McGaugh mass,\\   $\delta(log(M)) < 0.3$\end{tabular}               & $V_{75}$       & $3.48\pm0.03$          & $9.99\pm0.01$    & $0.004\pm0.003$           & $3.52^{+0.06}_{-0.05}$ & $9.96\pm0.01$       & $0.004\pm0.003$           \\
IIb  & \begin{tabular}[c]{@{}c@{}}Fiducial, Ponomareva mass, \\ $\delta(log(M)) < 0.3$\end{tabular}             & $V_{75}$       & $3.34\pm0.03$          & $9.77\pm0.01$    & $0.023^{+0.003}_{-0.007}$ & $3.62^{+0.07}_{-0.05}$ & $9.69\pm0.01$       & $0.000\pm0.00$            \\
IIc  & \begin{tabular}[c]{@{}c@{}}Fiducial, \\ recovered ``Taylor mass"\\    $\delta(log(M)) < 0.3$\end{tabular} & $V_{75}$       & $3.54\pm0.04$          & $9.93\pm0.01$    & $0.069\pm0.002$           & $3.51^{+0.06}_{-0.05}$ & $9.91\pm0.01$       & $0.03\pm0.01$             \\ \hline
IIIa & Fiducial                                                                                                 & $W_{50,P}$     & $3.35\pm0.04$          & $9.72\pm0.01$    & $0.08\pm0.01$             & $3.34\pm0.03$          & $9.66\pm0.01$       & $0.045\pm0.002$           \\
IIIb & \begin{tabular}[c]{@{}c@{}}Fiducial,\\  turbulence corrected $V_{75}$\end{tabular}                       & $V_{75,TC}$    & $3.22^{+0.04}_{-0.03}$ & $9.95\pm0.01$    & $0.07\pm0.002$            & $3.19\pm0.03$          & $9.90\pm0.01$       & $0.026\pm0.003$           \\
IIIc & Fiducial / Springob crossmatch                                                                           & $V_{75}$       & $3.33\pm0.07$          & $9.89\pm0.01$    & $0.032\pm0.003$           & $3.35^{+0.08}_{-0.07}$ & $9.87\pm0.02$       & $0.01\pm0.01$             \\
IIId & Fiducial / Springob crossmatch                                                                           & $W_{50,M}$     & $3.43^{+0.07}_{-0.06}$ & $9.52\pm0.02$    & $0.026\pm0.004$           & $3.44\pm0.06$          & $9.52\pm0.02$       & $0.02^{+0.004}_{-0.008}$  \\
IIIe & Fiducial / Springob crossmatch                                                                           & $W_{20,P}$     & $3.54^{+0.08}_{-0.07}$ & $9.41\pm0.02$    & $0.033\pm0.003$           & $3.56\pm0.07$          & $9.41\pm0.02$       & $0.02^{+0.005}_{-0.008}$  \\
IIIf & Fiducial / Springob crossmatch                                                                           & $W_{C}$        & $3.35\pm0.07$          & $9.69\pm0.02$    & $0.033\pm0.003$           & $3.37\pm0.06$          & $9.68\pm0.02$       & $0.007\pm0.005$           \\ \hline
IVa  & \begin{tabular}[c]{@{}c@{}}Fiducial / Springob crossmatch, \\ McGaugh Masses\end{tabular}                & $W_{C}$        & $3.66^{+0.08}_{-0.07}$ & $9.78\pm0.02$    & $0.029\pm0.004$           & $3.68\pm0.07$          & $9.78\pm0.02$       & $0.01\pm0.01$             \\
IVb  & \begin{tabular}[c]{@{}c@{}}Fiducial / Springob crossmatch, \\ Ponomareva masses\end{tabular}             & $W_{C}$        & $3.45\pm0.07$          & $9.73\pm0.02$    & $0.035\pm0.003$           & $3.45\pm0.07$          & $9.72\pm0.02$       & $0.02\pm0.01$             \\
IVc  & Gas fraction \textgreater 2                                                                              & $W_{50,P}$     & $3.80\pm0.08$          & $9.81\pm0.02$    & $0.089\pm0.002$           & $3.92^{+0.08}_{-0.10}$ & $9.77\pm0.02$       & $0.05\pm0.005$            \\
IVd  & $log(M_b) < 10.75 $                                                                                      & $W_{50,P}$     & $3.37\pm0.04$          & $9.74\pm0.01$    & $0.079\pm0.002$           & $3.39\pm0.03$          & $9.68\pm0.01$       & $0.046\pm0.001$           \\
IVe  & $b/a < 0.25$                                                                                             & $W_{50,P}$     & $3.34^{+0.07}_{-0.06}$ & $9.71\pm0.01$    & $0.066\pm0.002$           & $3.30\pm0.05$          & $9.70\pm0.01$       & $0.042\pm0.006$           \\ \hline
Va   & $log(M_b) > 10$                                                                                          & $V_{75}$       & $2.82\pm0.06$          & $10.02\pm0.02$   & $0.059\pm0.002$           & $2.88^{+0.06}_{-0.05}$ & $9.97\pm0.01$       & $0.011^{+0.007}_{-0.007}$ \\
Vb   & $log(M_b) > 10$                                                                                          & $V_{75,TC}$    & $2.78^{+0.05}_{-0.04}$ & $10.02\pm0.01$   & $0.060\pm0.02$            & $2.85\pm0.04$          & $9.97\pm0.01$       & $0.02^{+0.005}_{-0.01}$   \\
Vc   & $log(M_b) > 10$                                                                                          & $W_{50,P}$     & $2.83\pm0.07$          & $9.83\pm0.02$    & $0.071\pm0.002$           & $2.93^{+0.05}_{-0.05}$ & $9.75\pm0.02$       & $0.036\pm0.003$           \\
Vd   & $log(M_b) \leq 10$                                                                                       & $V_{75}$       & $4.13^{+0.12}_{-0.11}$ & $10.01\pm0.02$   & $0.066\pm0.002$           & $4.16^{+0.12}_{-0.11}$ & $10.00\pm0.01$      & $0.036\pm0.004$           \\
Ve   & $log(M_b) \leq 10$                                                                                       & $V_{75, TC}$   & $3.68\pm0.10$          & $10.00\pm0.01$   & $0.073\pm0.002$           & $3.77^{+0.10}_{-0.09}$ & $9.98\pm0.01$       & $0.040\pm0.004$           \\
Vf   & $log(M_b) \leq 10$                                                                                       & $W_{50,P}$     & $4.00^{+0.12}_{-0.11}$ & $9.73\pm0.01$    & $0.084\pm0.002$           & $4.08^{+0.12}_{-0.11}$ & $9.69\pm0.01$       & $0.055\pm0.002$           \\
Vg   & $D_{ALFALFA} \leq 40$Mpc                                                                                     & $V_{75}$       & $3.83^{+0.16}_{-0.15}$ & $9.92\pm0.04$    & $0.063\pm0.007$           & $3.86^{+0.17}_{-0.15}$ & $9.88\pm0.04$       & $0.03^{+0.01}_{-0.02}$    \\
Vh   & $D_{ALFALFA} \leq 70$Mpc                                                                                     & $V_{75}$       & $3.76^{+0.10}_{-0.09}$ & $9.94\pm0.02$    & $0.072\pm0.005$           & $3.67\pm0.08$          & $9.89\pm0.01$       & $0.03\pm0.01$             \\
Vi   & $D_{ALFALFA} \leq 100$Mpc                                                                                    & $V_{75}$       & $3.48\pm0.05$          & $9.92\pm0.01$    & $0.063\pm0.003$           & $3.43^{+0.05}_{-0.04}$ & $9.88\pm0.01$       & $0.025\pm0.05$            \\ \hline
VIa  & Fiducial, scatter = 0                                                                                    & $V_{75}$       & $3.15\pm0.02$          & $9.958\pm0.003$  & 0                         & $3.30^{+0.07}_{-0.03}$ & $9.90\pm0.01$       & 0                         \\
VIb  & Fiducial, scatter = 0                                                                                    & $W_{50,P}$     & $2.70\pm0.01$          & $9.81\pm0.002$   & 0                         & $3.28^{+0.05}_{-0.03}$ & $9.66\pm0.01$       & 0                        
\end{tabular}}
\caption{Summary of fits discussed in section~\ref{sec:discussion}. Horizontal bars demarcate different portions of the analysis - the first section demonstrates that relaxing the criteria which define $\alpha.R$ results in a relatively stable fiducial relation generally applicable to the full $\alpha.100$ sample. Note that we do not present results for a relation without relaxing restriction 2, which is that the sources must have stellar masses in our catalog, as it is not possible to place sources on our relation without a usable stellar mass estimate. The second region illustrates the differences between fits to the full fiducial sample using different stellar mass estimates, the third illustrates the difference induced by using different velocity definitions, the fourth provides relations with data methods / velocity definitions / sample criteria more closely matched to literature relations, and the fifth demonstrates the slope discrepancy between the high and low mass ends of the sample. \label{table:btfr_fits}}
\end{table*}

As mentioned in many previous comprehensive studies of the BTFR (e.g., \citealt{bradford16},~\citealt{ponomareva18}), factors including velocity width definition, sample mass range, fitting method, etc. must be considered when comparing BTFRs or using a BTFR as a template to minimize the biases introduced by a mismatch between the template-defining population and the sample population. The sample $\alpha.R$ used to derive the fiducial relation is a significantly restricted fraction of the full $\alpha.100$ sample, however as discussed below these restrictions improve the quality of the sample used to fit the template without significantly changing the best-fit relation, meaning that this template can be used on the full ALFALFA sample without introducing new biases. Conversely, comparing the relation presented here with other published relations illuminates the relative importance of some of these population characteristics on the best-fit relation. The comparisons presented in this section can be used to guide decision-making about whether or not it is appropriate to use this relation for a specific application.

\subsection{Consistency of results}\label{sec:consistent}

\begin{figure}
    \centering
    \includegraphics[width = \linewidth, trim = 75 75 75 75]{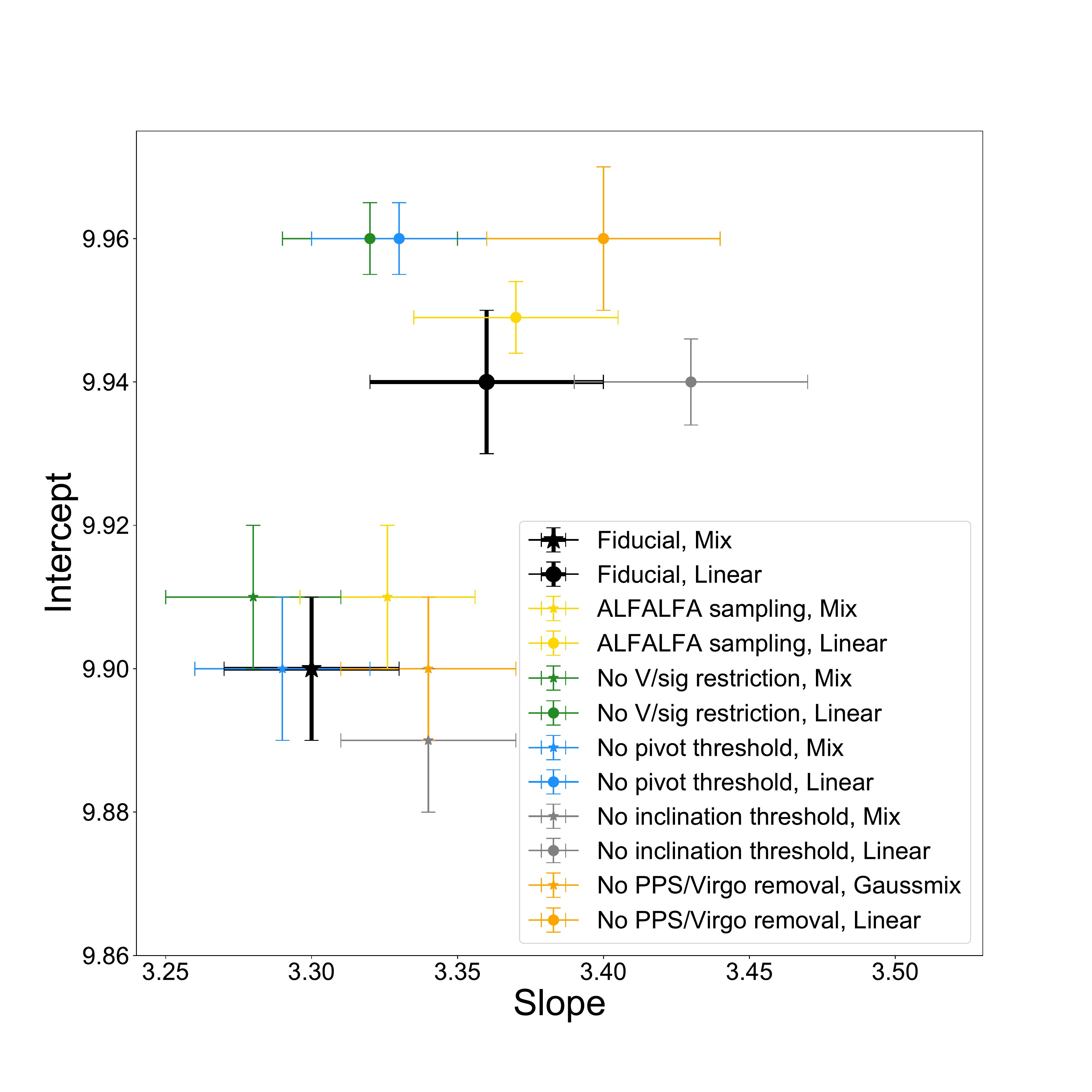}
    \caption{Summary figure showing consistency between fits to $\alpha.R$ and less-restricted subsamples of $\alpha.100$. The slopes and intercepts of all gaussian mixture models are consistent within $1\sigma$ (the uncertainties plotted here are the statistical uncertainties of the fitting procedure), indicating that the fiducial $\alpha.R$ BTFR is applicable for the larger, less restricted ALFALFA sample. The samples with all six selection criteria imposed ($\alpha.R$ and the ALFALFA-sampled sample) have the most consistency between the two models, indicating that the influence of outliers in the fit of these relations is effectively minimized by the selection criteria used.}
    \label{fig:consistency}
\end{figure}
To test the importance of each restriction used to define $\alpha.R$ on fiducial relation, we run additional fits, relaxing each of the criteria in sec.~\ref{sec:exclusions} in sequence. As demonstrated in Figure~\ref{fig:consistency}, the lines fit to less restricted versions of $\alpha.100$ are consistent with the fiducial relation. In all cases, the relation fit with an outlier component have higher intercepts than those fit without such a component, consistent with the fact that most of the outliers observed sit above and to the left of the bulk distribution of sources (see Fig.~\ref{fig:outliers}). As is shown by the positions of the black points in Figure~\ref{fig:consistency}, the fiducial ($\alpha.R$) sample has intercepts that are most similar between the simple and gaussian mixture models, indicating that the $\alpha.R$ sample contains fewer outliers which influence the best-fit relation than noisier samples which contain a larger portion of the $\alpha.100$ sample. 

\subsection{Reconciling differences with literature results}\label{sec:comparison}

The general agreement on a BTFR slope $3\lesssim m \lesssim 4$ does not converge to a narrower range across the literature, with slopes from 2.8 \citep{ponomareva18} to 4.1 \citep{papastergis16} reported for samples with differing characteristics including $V_{rot}$ definition \citep{lelli19}, the method for calculating $M_*$ \citep{ponomareva18}, and sample mass range \citep{karachentsev17}. We vary these parameters for subsamples of $\alpha.R$ where additional data is available in order to demonstrate how different measures of these parameters would affect the fiducial relation presented here.

\subsubsection{Stellar mass measurement method}\label{sec:stellarmass}

Even in high quality, restricted samples, galaxies' stellar masses often contribute significant uncertainty to the error budget of BTFR measurements. The details of stellar mass estimation methods vary from relation to relation, meaning that the stellar mass-to-light ratio, gas fractions, and baryonic mass measurements vary systematically between different relations, even when the same sample of galaxies is used. The fiducial relation presented here uses SDSS photometry-based stellar mass estimates which include a color correction, found to have generally best agreement with SED-based mass estimates over the ALFALFA sample \citep{durbala20}, and using a relation calibrated to maximize agreement with SED-based fits in the GAMA sample \citep{taylor11}.

Other recent relations prefer near-IR-based stellar mass estimates which commonly use a flat mass to light ratio ($\Upsilon_{3.6} \simeq 0.5$), as this gives the best consistency with optical stellar mass estimates \citep{mcgaugh14}, and also minimizes the vertical scatter of the best-fit BTFR \citep{lelli16b}. We fit a BTFR for the $\alpha.R$ sample using unWISE NIR photometry and this recommended $M/L$ ratio $0.5$, finding a significantly different slope than the fiducial relation (fit IIa). \citet{ponomareva18} use and advocate for a lower $M/L$ ratio of $0.36$, intermediate between dynamical and SED-based mass-to-light ratio estimates. Their recommended relation has a slope near the low end of literature values (2.8). Re-fitting the NIR-based relation using this lower $M/L$ ratio, we find a decrease in the slope of the relation from $3.48\pm0.03$ to $\sim 3.34\pm0.03$ (fit IIb), however the slopes from the $\alpha.R$ sample are still significantly higher than those for \citet{ponomareva18}, even after controlling for velocity definition, stellar mass estimating method, mass range and fitting method.

As illustrated by the differences between these fits and the fiducial relation, using different stellar mass estimation methods in an application sample will systematically offset sources from the fiducial relation, leading to biased distance estimates for that sample. However, we find that ``recovered” masses, which approximate SDSS-based stellar mass estimates using NIR photometric observations, can be used in the case of small samples to approximate the positions of sources on the fiducial relation. \citet{durbala20} present linear relations between multi-band GALEX-SDSS-WISE legacy catalog (GSWLC) and NIR and GSWLC and optical stellar mass estimates; combining these we have a simple linear prescription for removing first-order discrepancies between these estimates. We also fit a second-order function to a sample of sources that have both NIR and optical masses available in the \citet{durbala20} catalog. To determine the effect of these estimated stellar masses, we find the average residual for the $\sim1500$ remaining $\alpha.R$ sources not included in the fit using these ``recovered" masses. The average dispersion of these residuals is comparable to that of the optically-derived masses, suggesting that this may be a viable method for estimating stellar masses for small subsamples of an application sample where optical photometry is not available but NIR photometry is. However, we also check this consistency by fitting a relation to the 3000-galaxy $\alpha.R$ sample using the ``recovered" masses (fit IIc) and find a relation with a significantly steeper slope, potentially attributable to additional structure in the relationship between NIR-calculated stellar masses and the optical mass used here. \citet{kourkchi22} find a three-part piecewise in the NIR mass-to-light ratio derived to maximize agreement with optical masses, demonstrating the same higher-order structure in the relationship between mass estimates.

\subsubsection{Velocity definition}\label{sec:fit_veldef}

Comparisons between the BTFR presented here and those from other studies require the caveat that this relation uses a unique measure for rotational velocity. Relations fit for the same sample using different velocity measures generally have different slopes -- in particular, unresolved velocities and velocities measured using tracers which don't extend far into the galactic halo tend to give shallower slopes than tracers which reliably reach the extremes of the rotation curve (e.g., \citealt{brook16, ponomareva17,verheijen01}). 

The third section (e.g., fits IIIa-IIIf) of Table~\ref{table:btfr_fits} includes multiple relations fit to subsamples of $\alpha.R$ using different velocity definitions. As $\alpha.R$ is a subsample of the ALFALFA catalog, $W_{50,P}$ measurements are available for the full sample.  \citet{springob05}'s catalog of \HI linewidths includes widths $W_{20,P}$ (width at 20\% profile peak), $W_{50,M}$ (width at 50\% mean profile flux), and $W_C$ ($W_{50,P}$ corrected for broadening due to turbulence and instrumental effects) for 705 sources in the $\alpha.R$ sample. Though the $V_{75}$ relation for this subsample is consistent with the fiducial, the slopes of the $W_{20,P}$ and $W_{50,M}$ relations are marginally steeper ($\sim 1$ and $2\sigma$, respectively) and the intercepts are significantly lower. The higher slope of these relations suggests that these widths reach larger velocities at low mass and lower velocities at high mass -- this is consistent with the interpretation that these velocities may probe somewhat larger radii in the galaxy rotation curves \citep{lelli19}. The slope of the $W_C$ relation is most consistent with that of the $V_{75}$ fit relation, suggesting that $V_{75}$ requires less significant correction to account for turbulence and instrumental broadening than widths based on the peak of the line profile. A turbulence correction for $V_{75}$ is also proposed by \citet{yu20} -- though this is not included in this fiducial relation, we test this correction and find that the slope of the best-fit relation using $V_{75, TC}$ ($V_{75}$ with turbulence correction applied) decreases to $3.19\pm0.03$ (fit IIIb, Table~\ref{table:btfr_fits}). Such a decrease in slope is anticipated when subtracting a quantity which becomes a larger fraction of the width at small velocity widths. We do not include this correction in this first attempt at calibrating an ALFALFA BTFR with $V_{75}$ as it adds unnecessary complication to the relationship between data and application and does not notably improve the average uncertainty of the relation.

We find the least discrepancy when comparing literature relations to relations fit using the same velocity definitions on the ALFALFA data presented here. The relations fit to $\alpha.R$ using \citet{springob05} widths (IIId,e,f) have consistently lower slopes than those presented for the same velocity measures in \citet{lelli19}, however some of this discrepancy may be related to the different mass estimates -- using an IR-based mass, the best-fit slope for the $W_{50,M}$ \citet{springob05} crossmatch is $3.79\pm0.07$, consistent with the \citep{lelli19} slope ($3.62\pm0.09$) within $\lesssim 1.5\sigma$.Using the 41 galaxy overlap between the $\alpha.100$ sample and the SPARC sample \citep{lelli16}, we fit BTFRs using the masses, rotational velocities, and inclinations presented in both catalogs. Although the small sample size produces large uncertainties for best fit parameters, fits which use the $V_{75}$ velocity width measure and vary the inclination and stellar mass estimates are all consistent with the fiducial relation within $\sim1\sigma$. Fits which instead use a SPARC velocity definition are more consistent with the corresponding \citet{lelli19} relation (fit using an analogous likelihood function), again independent of the mass and inclination estimates used.

\subsubsection{Mass cutoff}
$\Lambda CDM$-based semianalytic models generally predict curved BTFRs \citep{trujillo-gomez11}, attributable to the decrease in baryon fraction at both high and low mass. Despite this, dedicated studies of the BTFR at low mass suggest the linear trend of the BTFR remains consistent to $M_{*} \sim 10^6 \ M_{\odot}$ \citep{iorio17}. Factors including still-rising rotation curves in dwarfs \citep{iorio17, catinella07} and the role of pressure- and rotation-support in low mass galaxies \citep{mcgaugh21} may affect the low mass end of the BTFR and are difficult to disentangle from observational effects. Possible non-linear structure in the BTFR means that the mass range covered by a BTFR template may affect the slope of the output relation. We test the magnitude of this effect by fitting BTFRs to high ($log(M_b) > 10$) and lower ($log(M_b) < 10$) mass subsamples of $\alpha.R$ using $V_{75}$, $W_{50,P}$, and $V_{75,TC}$ to discern the role of velocity definition and mass range in creating structure in these relations. As shown in fits Va-Vf of Table~\ref{table:btfr_fits}, the significantly ($\sim 9-10 \sigma$) different slopes of the high and low mass relations for all three velocity definitions point to additional, non-linear structure in this relation. Additionally, this supports the conclusion that best-fit relations are sensitive to the mass range of the sources probed. However, as discussed elsewhere (e.g.,\citealt{iorio17}), we caution overinterpretation of this variation, as only a comparison between a linear fit and a fit with physically-motivated curvature can truly distinguish these possibilities.

One important implication of this is that volume-limited samples are likely to have different best-fit relations, as these samples will include proportionately many more galaxies with lower masses. We run three fits with hard distance cuts at 40, 70, and 100 Mpc to determine the magnitude of this effect and find, as expected, that the sample with a distance cut of 40 Mpc has a significantly ($\sim3.7\sigma$, fit Vg) higher slope than both the fiducial relation and the samples with 70 and 100 Mpc distance limits (fits Vh and Vi, respectively). The 3$\sigma$ difference for the 100 Mpc cut is somewhat less significant -- once the mass function is well-sampled at all masses, the density of points along the relation is a secondary effect in determining the slope. To confirm this, we also run fits which force quasi-uniform sampling, drawing 9 galaxies per 0.25 dex mass bin between 7.5-11.5 log($M_b$). With only 146 galaxies per fit, these relations are sensitive to the sample drawn (consistent with \citet{sorce16}), however they give on average fits that have a somewhat steeper slope ($\sim 3.5$ vs. 3.3) than the fiducial relation.

\subsubsection{Gas Fraction}

Though \citet{papastergis16} also uses a subsample of ALFALFA to fit their relation, their BTFR has a drastically different slope of $4.15\pm0.23$. The significant demographic differences between this sample and $\alpha.R$ are: (1) A stringent quality threshold, including both a S/N cut and a manual inspection of all candidates above this threshold, (2) a larger axial ratio cut of b/a $<$ 0.25 (corresponding to an inclination cut $i > 78^\circ$), and (3) a stringent cut to include only gas-rich sources in the sample, with $M_{HI}/M_{*} > 2$. To determine which of the above factors is influential in this discrepancy, we run fits to subsamples of $\alpha.R$ using these cuts (see fits IVc and IVd in Table~\ref{table:btfr_fits}). Additionally, low mass galaxies are more likely to be gas-rich, so restriction (3) changes the distribution of masses of the \citet{papastergis16} sample to include fewer high mass sources, with no sources with $M_b > 10^{11} M_\odot$. To differentiate between the effects of changing the mass range and the gas fraction distribution, we also run a fit on a subsample of $\alpha.R$ with an upper mass limit of $M_b < 10^{10.75} M_\odot $ (fit IVe). 

The best fit slope for the sample with a gas fraction restriction imposed is $3.92^{+0.08}_{-0.10}$, significantly higher than the other slopes for $\alpha.R$ subsamples. \citet{lelli16b} similarly find that weighting sources to prefer gas-rich galaxies marginally increases the slope of their best-fit BTFR. These results are consistent with the finding of \citet{huang12} that, at a given mass, galaxies with a higher spin parameter have a higher gas fraction. Spin parameter is inversely proportional to rotational velocity, meaning {\em a flat cut in gas fraction preferentially selects for slower rotators at high mass, increasing the slope of the relation}. In this, the primary difference between the slope of the fiducial relation and \citet{papastergis16} is the flat gas fraction cut, making this another parameter that should be consistent between application and template samples.

\subsubsection{Fitting Routine}
In general, astrophysical problems with errors in both the dependent and independent variables require rigorous treatments of model-fitting \citep{hogg10}. For consistency in the above comparisons, we run data from literature results through our fitting routine, as previous studies have discussed and demonstrated \citep{bradford16} how this effect can impact the BTFR. The fitting routine chosen here is used to minimize the average orthogonal distance of a point, weighted by both the x and y uncertainties, from the BTFR. We prefer this over a relation that looks to minimize scatter in the y direction alone, as this takes into account the significant uncertainty in the x variable (inclination-corrected rotational velocity). This can present an additional complication in comparing the fiducial relation here with other studies -- for example, though the $W_{20,P}$ relation presented here ($m = 3.54^{+0.08}_{-0.07}$) is steeper than the fiducial \citet{bradford16} relation ($ m = 3.24\pm 0.05$), it is in good agreement with the relation fit to their sample using the orthogonal SIXLIN algorithm ($m = 3.6$).

In addition, these relations include an intrinsic scatter component which is part of the best fit. This allows the relation to have a non-zero perpendicular width, which is motivated in $\Lambda$CDM due to the non-negligible distribution of halo spins and baryon fractions and may, in observational samples, also be partially attributable to underestimated uncertainties (e.g., like the underestimated inclination uncertainties discussed previously). However, this also affects the best-fit relations, as non-linear structure including curvature may be fit as part of the non-zero scatter rather than as a feature in the shape of the relation. We find, consistent with \citet{papastergis16}, that relations without an intrinsic scatter component have shallower slopes (see, e.g., the differences between Ia vs. VIa, IIIa vs. VIb in Table~\ref{table:btfr_fits}), and that this effect roughly scales with the size of the best fit scatter (which is clearly visible in the smaller discrepancy between the best-fit mixture models with/without scatter). These results are in agreement with previous work by \citet{weiner06} on the TFR, further demonstrating the role of the fitting routine in determining the presented relation.

\section{Application: Recovering Virgo Infall}
We use the BTFR presented here to find distances to sources near Virgo, recovering signatures of infall onto the nearby ($\sim 17$ Mpc) cluster. Previous studies using large samples of galaxies with relatively precise (5-25\% uncertainties) distance measurements \citep{karachentsev10} and high-precision distances measured using the tip of the red giant branch in resolved color magnitude diagrams \citep{karachentsev18}, find clear signatures of infall attributable to a mass overdensity of $\sim 8.6\times10^{14} M_\odot$. 

Within 15 degrees of M87 (RA 12h30m49.42s Dec +12d23m28.043s; here considered the center of Virgo), there are 517 sources in the ALFALFA catalog with $V_{hel} < 3300$ km/s. Of these sources, 154 are in the subsample of ALFALFA including all $\alpha.R$ selection criteria but \#3 ({\em Sources far from large overdensities}) -- we test best-fit infall using this sample (called ``all Virgo vicinity, $\alpha.R$" in Table~\ref{table:infall_fits}), as well as a subsample of sources in the Virgo vicinity unlikely to be members of the virialized cluster core by removing Extended Virgo Cluster Catalog members \citep{kim14}, leaving 80 sources in the ``pure infall"/``No EVCC members, $\alpha.R$" 15 degrees sample. 

In a Virgocentric reference frame, we expect source infall on a single-attractor to follow the expression discussed in \citet{karachentsev18, karachentsev10}:
\begin{equation}\label{eqn:infall}
\begin{split}
        V_{in}(r_{Virgo}) = &H_0\times(r_{Virgo}) - \\ & H_0\times r_0 \times \sqrt{\frac{r_0}{r_{Virgo}}}
\end{split}
\end{equation}

Where $r_{Virgo}$ is the distance of the source to Virgo 
\begin{equation}
\begin{split}
& r_{Virgo} = \\ & \sqrt{D_{Gal}^2 + (D_{C} + D_{C, off})^2 - 2D_{Gal}(D_{C} + D_{C, off})\cos\Theta}
\end{split}
\end{equation}
The infall onto Virgo can thus be described by the parameters $r_0$ (the zero-velocity radius of the cluster) $H_0$ (Hubble parameter, may depart significantly from 70 km/s in this model), $V_{C, off}$ (offset of the cluster velocity from fiducial 984 km/s in LG-centric frame), $D_{C, off}$ (offset of cluster center from fiducial 17 Mpc distance), and $V_{err}$, the scatter in velocities due to random motion. Note that $\Theta$ is the angular separation of the galaxy from the cluster center (M87) and that $D_{Gal}$ is the distance to the galaxy from the observer.

\begin{figure*}
    \centering
    \includegraphics[width = 0.45\linewidth]{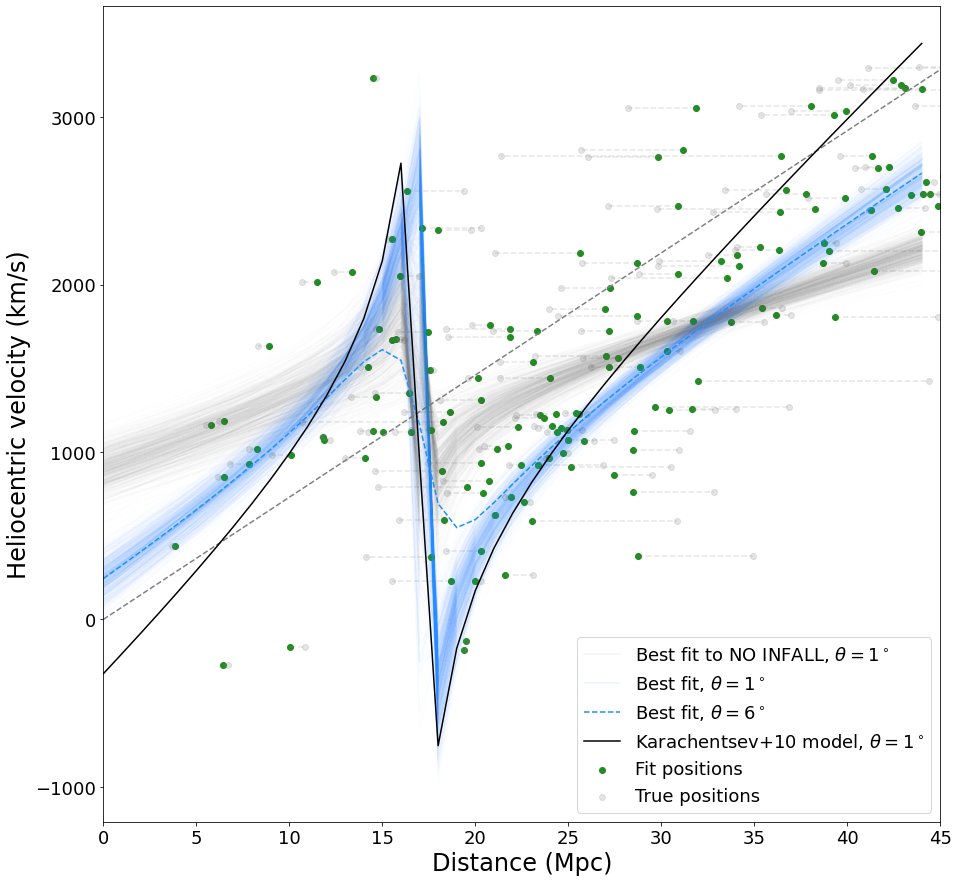}
    \includegraphics[width = 0.45\linewidth]{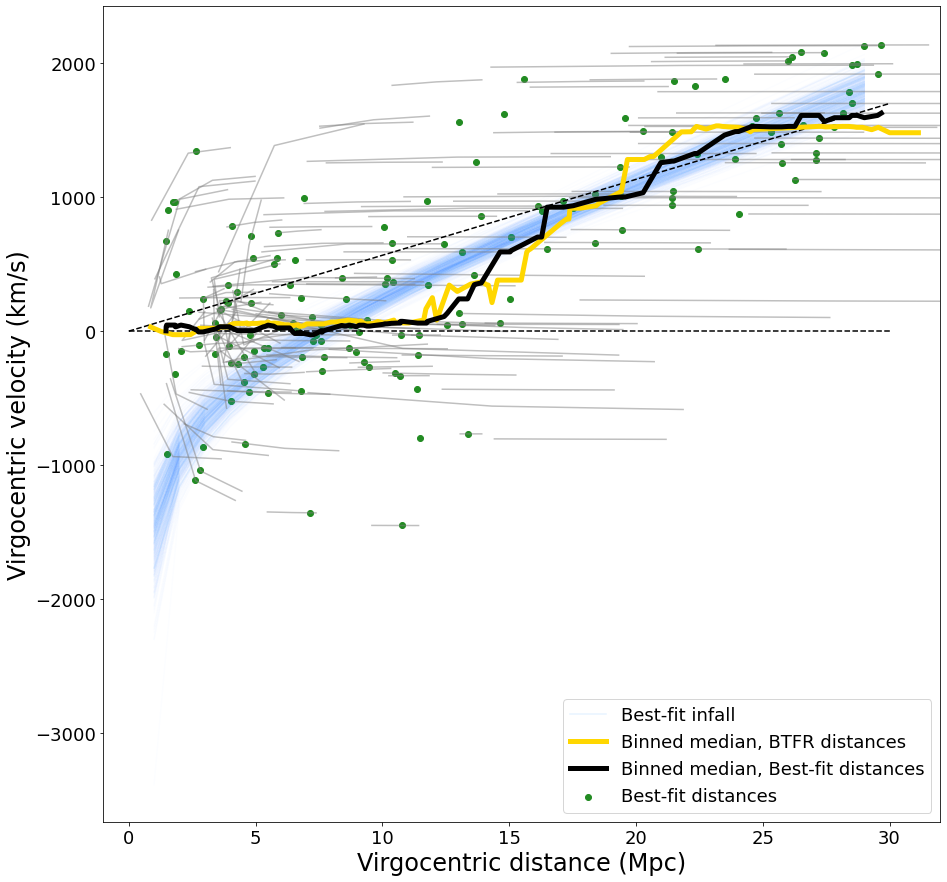}
    \caption{Infall fits to $\alpha.R$ sources within $\Theta < 15^\circ$ of Virgo. Left: fit and data in heliocentric reference frame, using $D_{BTFR}$ and $V_{LG}$. The best-fit infall model from \citet{karachentsev10} is plotted in black, with the best-fit model from this analysis overplotted in blue (solid vs. dashed lines indicate the magnitude of infall at different angular radii from the cluster; infall at 1$^\circ$ is plotted using 600 random samples from the posterior). The green points in this figure are the best-fit distances to the galaxies in the sample, which are recovered from the likelihood's posterior. These distances are connected to each source's corresponding BTFR distance with faint, dashed grey lines. The grey traces are drawn from the posterior of an analogous fit for a cluster-free sample (Non-Virgo, $\alpha.R$). Right: Fit and sample data in Virgocentric reference frame, with best-fit infall overplotted in blue in comparison with pure Hubble flow in the vicinity of the cluster (grey), as well as binned median Virgocentric velocities using the measured best-fit distances to the galaxies ($D_{BTFR}$; gold solid line) and the average best-fit distances from the Bayesian model (black solid line) -- the green points are the best-fit distances, while the grey lines near each point give the BTFR distance to each source $\pm$ the uncertainty, indicating how the positions and Virgocentric distances of the sources change as the sources distances are fit.}
    \label{fig:virgo_infall}
\end{figure*}

These distances can be measured using the fiducial BTFR (eq.~\ref{eqn:BTFR}), $V_{75}$ measurements, \HI fluxes, and SDSS source magnitudes. We find good agreement (mean difference = 0, dispersion 1$\sigma$; see Figure~\ref{fig:distance_comparisons}) within the reported uncertainties for 109 sources with independent distance measurements \citep{kashibadze20}, underscoring the importance of properly handling uncertainties when modeling infall. With these distances, we plot binned medians of the Virgocentric infall of these galaxies in Figure~\ref{fig:virgo_infall}, along with the best-fit model (eqn.~\ref{eqn:infall}). Though these curves appear consistent with previous studies \citep{karachentsev10} and with expectation (near 0 within the virial radius of the cluster, decreasing below Hubble flow outside of the virial radius, increasing toward Hubble flow at larger distances), the relatively small magnitude of this velocity difference (few 100 km/s) is difficult to see in the raw data, and likely to have smaller magnitudes in binned averages given the significant distance uncertainties in the data.

Model fitting provides another avenue for describing this signature of infall. To handle the significant distance uncertainties, the model used here simultaneously fits for the true distance to each galaxy within the sample as well as parameters describing the infall onto the cluster. This large number of free parameters necessitates the use of an efficient method for sampling this likelihood function. We use pySTAN, a Python implementation of the Stan Hamiltonian Monte Carlo statistical analysis program\footnote{Stan Development Team, 2021. Stan Modeling Language Users Guide and Reference Manual, 2.28. https://mc-stan.org} to evaluate this likelihood. 

\begin{table*}[]
\centering

\begin{tabular}{llllll}
\hline
Sample                                                          & $r_0$         & $H_0$        & $D_{off}$     & $V_{C,off}$    & $\sigma_V$     \\ \hline
All Virgo vicinity, $\alpha.R$                                  & $7.58\pm0.45$ & $72.0\pm4.0$ & $0.25\pm0.23$ & $95.3\pm41.9$  & $351.7\pm15.6$ \\
No EVCC members, $\alpha.R$                                     & $7.8\pm0.7$   & $83.7\pm5.8$ & $0.05\pm0.25$ & $145.8\pm60.3$ & $290.6\pm18.2$ \\
Non-Virgo, $\alpha.R$                                           & $4.3\pm1.9$   & $29.5\pm3.0$ & $0.01\pm0.00$ & $505.4\pm37.3$ & $453.8\pm11.8$  \\
\end{tabular}
\caption{Table detailing infall fits to sources in the vicinity of Virgo, including excision of cluster members via the EVCC \citep{kim14}, as well as a comparison sample away from Virgo \label{table:infall_fits}}
\end{table*}

We assume a $\propto d^2/D_{max}^3$ prior on the distance to the galaxies, allowing the sources to preferentially be further than their measured distances. Such a prior corrects for the Malmquist bias of a flux-limited sample, which is roughly applicable for our sample. We use gaussian priors for the remaining parameters: $\mu_{r0},\sigma_{r0} = 7.3, 3.7$ Mpc,  $\mu_{H0},\sigma_{H0} = 73, 50$ km/s/Mpc,  $\mu_{D_{C,off}}, \sigma_{D_{C,off}} = 0, 0.25$ Mpc,  $\mu_{V_{C,off}}, \sigma_{V_{C,off}} = 0, 100$ km/s,  $\mu_{Verr}, \sigma_{Verr} = 100, 25$ km/s. These priors are also truncated to avoid the unphysical situations of negative dispersion, $r_0$, and $H_0$ values. The priors on $D_{C,off}$ and $V_{C,off}$ are slightly more constraining in order to avoid one common poor fit, wherein the S-shape (see the blue and black curves in Figure~\ref{fig:virgo_infall}) can move to a poorly-sampled region of distance-velocity space which is unlikely to be the true position/velocity of the cluster.

A summary of these fits is collected in Table~\ref{table:infall_fits}. In the case of the fits run on Virgo vicinity samples, the highly non-zero values of $r_0$, sensible values of $H_0$ and the position and velocity of the cluster suggest that we are able to recover the signature of infall from this data. In order to provide a ``null result" comparison, we also run this fit on a sample of sources selected to be $\Theta >> 15^\circ$ from Virgo (``Non-Virgo, $\alpha.R$" in Table~\ref{table:infall_fits}). As is seen in the third row of this table, this best-fit model finds a smaller, less significantly non-zero $r_0$ (we expect $\rightarrow 0$ in the zero infall case) as well as a low $H_0$ and large $V_{c, off}$ value. We note that the best-fit $H_{0}$ for the ``null" sample is significantly lower than the true value, which is not consistent with the expectation that this relation should converge to Hubble flow in the presence of zero infall. Instead, this result moves the sharpest features of the relation to a poorly-constrained portion of distance-velocity space (above \& in front of simulated ``Virgo", $V_{hel} > 1000$ km/s, $D_{gal} < 15$ Mpc), and the preferentially larger uncertainties and distance priors allow this to occur by placing distances further ``behind" the simulated cluster, decreasing the best-fit $H_0$. This demonstrates that priors which broadly reflect the expected cluster position and velocity are important in this fitting process.

\section{Conclusions}
Using a velocity definition previously unused in BTFR applications ($V_{75}$, where the integrated flux of a line profile reaches 75\% of the ``flat, fully integrated flux"), we present a BTFR representative of the full ALFALFA sample:\begin{equation}
    \log(M_{b}/M_\odot) = (3.3\pm0.06)(\log(V_{rot,75})-2)+(9.9\pm0.01)
\end{equation}
where $V_{rot,75}$ is $V_{75}$ corrected for inclination and redshift. We fit this fiducial relation with a highly-restricted portion of the full $\alpha.100$ sample ($\alpha.R$, a sample of $\sim4500$ galaxies or $\sim 10\%$ of $\alpha.100$) and demonstrate that the full ALFALFA sample agrees with this relation, albeit with an expected increase in uncertainty and outlier fraction. This relation is fit using a gaussian mixture model to simultaneously find the best-fit BTFR while controlling for the fraction of unreliable but difficult to excise data (a common issue for large survey datasets). We demonstrate that the majority of the sources downweighted by this process are justifiably separate from the reliable measurements and regularly-rotating spiral galaxies that define the bulk of the $\alpha.R$ relation. An overwhelming majority of these sources have reasons for being offset from the relation including a close companion, an overestimated inclination with underestimated uncertainties, a stellar mass that may be unreliable, or a velocity width that may be unreliable. Altogether, these account for $\sim 70\%$ of the most discrepant outliers, motivating our use of a mixture model to downweight these sources which are difficult to identify automatically. 

Using subsets of the fiducial ($\sim 4500$ galaxy) sample, we demonstrate the effects of mass sampling, stellar mass estimation methods, velocity width measurement methods, and various selection effects on the slope of the best-fit BTFR, collecting these in Table~\ref{table:btfr_fits}. Comparing the fiducial relation with these fits motivates us to make recommendations for circumstances where the relation presented here should/should not be applied.

The following is a summary of recommendations for use of the BTFRs presented in this paper:

\begin{enumerate}
\item Application samples should use the same velocity definition used in the fitting sample. As noted in other detailed studies of the BTFR, we find marginal differences in slope when using different velocity definitions to determine $V_{rot}$, even though all of the definitions used here are measures derived from the unresolved line profile. 

\item Samples which have significantly different mass ranges, or which have drastically different mass samplings should not be used with this relation. Forcing high or low mass samples created the largest discrepancies of the tested demographic cuts, indicating the presence of additional non-linear structure independent of the velocity definition used to calibrate this relation. If this relation is to be applied to volume-limited samples, the mass range of those samples will determine to first-order their agreement with the presented relation. Forcing uniform sampling has a marginal effect on the relation ($\sim2\sigma$ slope difference), consistent with the expectation that a sensitivity-limited sample oversamples the most massive, brightest galaxies.

\item Samples should use consistent stellar mass estimation methods -- in this case, optical photometry-based estimates of galaxy stellar masses. For small subsamples where no such data is available, corrected NIR estimates can be used, however using these estimates for the full sample results in a significantly different slope to the best-fit relation.

\item Samples should use matched flat gas fraction cuts. We find that imposing a flat gas fraction cut preferentially includes galaxies at lower $V_{rot}$ at log($M_b) >$ 9, increasing the slope of the relation.  
\end{enumerate}
Within the range $3-4$, an ALFALFA-based BTFR will have a slope largely dictated by the properties of the sample used to calibrate the relation. Through cherry-picked subsamples of $\alpha.R$ and the use of different width and stellar mass definitions, we found best-fit slopes across the range $2.8\sim4.1$. This has implications for the use of the BTFR as a distance indicator, as the sample the template is applied to should have similar properties (mass range, mass sampling, gas fraction, velocity definition, stellar mass definition) to the sample used to calibrate the relation. Additionally, this demonstrates the challenge of observationally determining the slope, scatter, and thus cosmological significance of the ``fundamental" BTFR. 

The relation presented here has application to HI-selected surveys, and is particularly optimized for large samples which may include spurious data points. The curve of growth-based velocity definition used here, first presented in \citet{yu20}, is similarly ideal for use in large surveys as it can be calculated completely automatically (as has been done with related velocity width $V_{85}$ for the full ALFALFA sample \citep{yu22a}). Through multiple tests, we demonstrate that the returned widths and their uncertainties are generally successful descriptions of known velocity widths, even when profiles have been degraded to S/N comparable with average sources in the ALFALFA sample and when they include simulated baseline effects. We estimate $< 7\%$ of sources in a sample with an ALFALFA-like S/N distribution have underestimated uncertainties, based on the success of the width-fitting program at recovering these known velocities within $3\sigma$.

In conclusion, this work demonstrates that large surveys which can be efficiently and uniformly reduced using automated methods provide a test for how different sample properties or selection effects may bias a resulting BTFR. Future large-scale \HI surveys including CRAFTS, undertaken by FAST \citep{zhang21}, and the recent MIGHTEE-HI survey undertaken with MeerKAT \citep{ponomareva21} will continue to increase our sensitivity to HI-rich galaxies at lower masses and larger distance and redshifts than before, meaning that the BTFR will continue to be a compelling tool for measuring distances and galaxy evolution over cosmic time.

\begin{acknowledgements}
In memory of Riccardo Giovanelli, without whom this work would not exist, and Christopher Springob, whose research was foundational to much of our work. CJB and MPH acknowledge support from NSF/AST-1714828 and grants from the Brinson Foundation. We thank all members of the Undergraduate ALFALFA Team for their contributions, especially Mary Crone-Odekon, Aileen O'Donoghue, Anhad Gande, and Rose Finn, as well as NSF grants AST-1211005, AST-1637339, and AST-2045369 for support. We acknowledge the usage of the HyperLeda database (http://leda.univ-lyon1.fr). This research made use of Astropy\footnote{http://www.astropy.org}, a community-developed core Python package for Astronomy \citep{astropy13, astropy18} and matplotlib \citep{hunter07}.


From 2011 to 2018, the Arecibo Observatory was operated by SRI International under a cooperative agreement with the National Science Foundation (AST-1100968) and in alliance with Ana G. Méndez-Universidad Metropolitana, and the Universities Space Research Association. Currently, the Arecibo Observatory is a facility of the National Science Foundation operated under cooperative agreement (AST-1744119) by the University of Central Florida in alliance with Universidad Ana G. Méndez (UAGM) and Yang Enterprises (YEI), Inc.
{\em Facilities:} Arecibo \\
{\em Software:} AstroPy \citep{astropy13,astropy18}, emcee \citep{foremanmackey13}, Matplotlib \citep{hunter07}, NumPy \citep{harris20}, pySTAN
\end{acknowledgements}

\appendix

\section{BTFR Fitting}
As discussed elsewhere, using various methods for fitting relations between variables with uncertainty in both x- and y- variables can affect output BTFRs (e.g., \citet{bradford16}). For reproducibility, we provide the derivation of the likelihood used here. As used in \citet{lelli19},\citet{papastergis16}, and discussed in \citet{hogg10}, this likelihood is based on minimizing the scaled orthogonal distance of data points from the central relation. As is regularly done in studies of the BTFR, we use a relation that simultaneously fits for the slope, intercept, {\em and intrinsic scatter} of the relation -- in the case of this likelihood, the intrinsic scatter is a gaussian convolved perpendicular to the best-fit line. Considering $\theta = \arctan{(m)}$, $\phi = \arctan{(\frac{y-b}{x})}$, we find $d_\perp = D\sin{(\theta - \phi)}$, where $D = \sqrt{x^2 + (y-b)^2}$. Rewriting,
\begin{equation}
    d_\perp = \sqrt{x^2 + (y-b)^2}\bigg(\sin\theta\cos\phi - \cos\theta\sin\phi\bigg)
\end{equation}
$\sin\phi = \frac{y-b}{x}\frac{1}{\sqrt{1+(\frac{y-b}{x})^2}} = \frac{y-b}{\sqrt{x^2+(y-b)^2}}$
and $\cos\phi = \frac{1}{\sqrt{1+(\frac{y-b}{x})^2}} = \frac{x}{\sqrt{x^2+(y-b)^2}}$ gives:
\begin{equation}
    d_\perp = \sin\theta x - \cos\theta (y - b) = \sin\theta x - \cos\theta y + b_\perp
\end{equation}.
Again, if data is distributed about the best-fit relation with an intrinsic Gaussian scatter $\sigma_\perp$, then the probability of a set of data parameters $(\Tilde{m},\Tilde{v})$ given model parameters $(\theta, b_\perp, \sigma_\perp)$:
\begin{equation}
    p(\Tilde{m},\Tilde{v}|\theta, b_\perp, \sigma_\perp) = \frac{1}{\sqrt{2\pi\sigma^2_\perp}}\mathrm{exp}\bigg[-\frac{(\sin\theta\Tilde{v} - \cos\theta\Tilde{m} + b_\perp)^2}{2\sigma^2_\perp}\bigg]
\end{equation}
However, we don't observe $\Tilde{m},\Tilde{v}$ -- instead, we observe $\Tilde{m}_o,\Tilde{w}_o,$ and $\Tilde{s}_o$, where $\Tilde{v} = \Tilde{w} - \Tilde{s}$, and the Tilde denotes the log of the observed quantity (baryonic mass, velocity width, and sin($i$)). In theory, we want to marginalize over all possible values of the true $\Tilde{m},\Tilde{w},\Tilde{s}$ values:
\begin{equation}
    p(\Tilde{m}_o,\Tilde{w}_o,\Tilde{s}_o |\theta, b_\perp, \sigma_\perp) = 
    \int\int\int d\Tilde{m}d\Tilde{w}d\Tilde{s} p(m)p(w)p(s) \frac{1}{\sqrt{2\pi\sigma^2_\perp}}\mathrm{exp}\bigg[-\frac{(\sin\theta\Tilde{v} - \cos\theta\Tilde{m} + b_\perp)^2}{2\sigma^2_\perp}\bigg]
\end{equation}
If we assume that the errors in all three variables are Gaussian distributed, then we can use the fact that a convolution of two Gaussians is itself a Gaussian to marginalize as desired, giving the likelihood:
\begin{equation}
    p(\Tilde{m}_o,\Tilde{w}_o,\Tilde{s}_o |\theta, b_\perp, \sigma_\perp) = 
     \frac{1}{\sqrt{2\pi(\sigma^2_\perp+\sin^2\theta(\sigma_{w,o}^2+\sigma_{s,o}^2) + \cos^2\theta\sigma^2_{m,o}}}\mathrm{exp}\bigg[-\frac{(\sin\theta(\Tilde{w_o} -\Tilde{s_o}) - \cos\theta\Tilde{m_o} + b_\perp)^2}{2(\sigma^2_\perp+\sin^2\theta(\sigma_{w,o}^2+\sigma_{s,o}^2) + \cos^2\theta\sigma^2_{m,o})}\bigg]
\end{equation}
In reality, likelihoods can be built to more correctly address the asymmetries of these errors due to their true asymmetry / non-gaussianity, however depending on the expression of the errors chosen, these integrals do not marginalize out and the computational requirements for evaluating the likelihood function become severe. Using pySTAN, we test the effect of relaxing some of these criteria, and do not find that the improvements in accuracy of this fit scale with the increasing complexity of these models, especially given selection criteria which select for the largest uncertainties and thus the datapoints where these effects will be most significant.



\bibliography{mybib}

\end{document}